\documentclass[reprint,aps,pra,showpacs,superscriptaddress,longbibliography]{revtex4-1}
\pdfoutput=1
\usepackage{amsmath,amssymb,graphicx,dcolumn,hyperref,graphicx}
\usepackage{hyperref}
\usepackage{epstopdf}
\hypersetup{colorlinks=true,linkcolor=blue,citecolor=blue,urlcolor=blue}
\usepackage{mathptmx}

\usepackage[below]{placeins}
\usepackage{xcolor}
\definecolor{bblue}{HTML}{4b72b0}
\definecolor{rred}{HTML}{c55055}
\definecolor{ggreen}{HTML}{58a96a}

\renewcommand{\vec}[1]{\mathbf{#1}}
\newcommand{\eb}{\varepsilon_b}
\newcommand{\eo}{\varepsilon_0}
\newcommand{\Hbar}{\overline{\text{H}}}

\let\originalleft\left
\let\originalright\right
\renewcommand{\left}{\mathopen{}\mathclose\bgroup\originalleft}
\renewcommand{\right}{\aftergroup\egroup\originalright}

\begin{document}
\frenchspacing

\title{Formation of positron-atom bound states
in collisions
between Rydberg Ps and neutral atoms}
\author{A. R. Swann}\email{aswann02@qub.ac.uk}
\affiliation{School of Mathematics and Physics, Queen's University Belfast, University Road, Belfast BT7 1NN, United Kingdom}
\author{D. B. Cassidy}\email{d.cassidy@ucl.ac.uk}
\affiliation{Department of Physics and Astronomy, University College London, Gower Street, London WC1E 6BT, United Kingdom}
\author{A. Deller}\email{a.deller@ucl.ac.uk}
\affiliation{Department of Physics and Astronomy, University College London, Gower Street, London WC1E 6BT, United Kingdom}
\author{G. F. Gribakin}\email{g.gribakin@qub.ac.uk}
\affiliation{School of Mathematics and Physics, Queen's University Belfast, University Road, Belfast BT7 1NN, United Kingdom}
\date{\today}

\begin{abstract}
Predicted twenty years ago, positron binding to neutral atoms has not yet been observed experimentally. A new scheme is proposed to detect positron-atom bound states by colliding Rydberg positronium (Ps) with neutral atoms. Estimates of the charge-transfer reaction cross section are obtained using the first Born approximation for a selection of neutral atom targets and a wide range of incident Ps energies and principal quantum numbers.
We also estimate the corresponding Ps ionization cross section.
The accuracy of the calculations is tested by comparison with earlier predictions for Ps charge transfer in collisions with hydrogen and antihydrogen. We describe an existing Rydberg Ps beam suitable for producing positron-atom bound states and estimate signal rates based on the calculated cross sections and realistic experimental parameters. We conclude that the proposed methodology is capable of producing such states and of testing theoretical predictions of their binding energies. 
\end{abstract}


\maketitle

\section{\label{sec:intro}Introduction}

Being antimatter particles, positrons ($e^+$) are of fundamental importance for tests of QED and the Standard Model~\cite{Karshenboim2005,Ishida14,Schael06}, and in astrophysics~\cite{Guessoum14}. They also find numerous applications in condensed matter physics, surface science, atomic physics, and medicine (see, e.g., Refs.~\cite{Tuomisto13,Wahl02,Charlton2001}). Though positrons were discovered more than eighty years ago~\cite{Anderson33}, there is still much about their interactions with matter that is not fully understood.

One such outstanding question is positron binding to neutral atoms.
Positron-atom bound states were first predicted by many-body-theory calculations in 1995~\cite{Dzuba95}. Two years later, variational calculations carried out by Ryzhikh and Mitroy~\cite{Ryzhikh97} and Strasburger and Chojnacki~\cite{Strasburger98} confirmed that a positron can bind to lithium. Soon after, many calculations of positron binding to other atoms appeared; see Ref.~\cite{Mitroy02} for a 2002 review. Despite a wealth of predictions for positron-atom binding energies now available (for a survey of the Periodic Table, see Refs.~\cite{Dzuba12,Harabati14}), no experimental evidence of positron-atom bound states has yet arisen. This is chiefly due to the limited availability of suitable positron sources, the difficulty in obtaining the required neutral atom species in the gas phase, and the need to implement an efficient production and unambiguous detection schemes.

The situation for positron binding with molecules is essentially the opposite
\cite{Gribakin10}. Positron annihilation in polyatomic molecules is typically mediated by positron capture in vibrational Feshbach resonances (VFR), where the positron enters a quasibound state by transferring its excess energy into molecular vibrations of a single mode with near-resonant energy. By using a trap-based positron beam~\cite{Gilbert97,Kurz98}, experimentalists were
able to observe VFRs in the positron energy dependence of the annihilation rate~\cite{Gilbert02}. The downshift of a resonance relative to the vibrational excitation energy provided a measure of the positron binding energy. This has enabled positron binding energies to be determined for over seventy molecules~\cite{Danielson09,Danielson10,Jones12,Danielson12}. On the side of theory, there are few calculations of positron binding to nonpolar or weakly polar molecules. The zero-range potential model~\cite{Gribakin06,Gribakin09} captured the qualitative features for the alkanes, and there were configuration-interaction (CI) calculations for carbon-containing triatomic molecules~\cite{Koyanagi13,Koyanagi13a}. For strongly polar molecules many quantum-chemistry calculations have been performed, but only a few of them allow direct comparison with experiment; recent CI calculations for nitriles, aldehydes, and acetone~\cite{Tachikawa11,Tachikawa12,Tachikawa14} gave binding energies within 25--50\% of experimental values. A simple theoretical model was recently proposed to explain the dependence of the binding energy on the molecular dipole moment and dipole polarizability~\cite{Gribakin15}.

Regarding positron-atom bound states, several ways of detecting them in experiment have been proposed. In Ref.~\cite{Mitroy99} it was suggested that positronic atoms could be formed in collisions with negative ions, $e^++A^-\rightarrow e^+A+e^-$,
the positron affinity determining the energy threshold of this reaction.
As is the case of molecules, for some atoms it may be possible to observe resonances in the positron annihilation rate and associate these with binding~\cite{Dzuba10}. Another scheme for measuring positron-atom binding energies is laser-assisted photorecombination of positrons from a trap-based beam with metal atoms in a vapor~\cite{Surko12}. It may also be possible to capture positrons into shallow bound levels using pulses of a very strong magnetic field~\cite{Harabati14}.

Here we propose an alternative strategy for the creation and detection of positron-atom bound states in charge-exchange collisions of Rydberg-state positronium (Ps) with neutral atoms. Rydberg Ps was first generated by Ziock \textit{et al.} using a linac-based positron beam~\cite{Ziock90}, but it was only possible to demonstrate the production of a few high-lying states  with principal quantum numbers $n=13$--15. Modern positron-trapping~\cite{Danielson15} and detection~\cite{Cooper2015} techniques have facilitated much more efficient production of Rydberg Ps~\cite{Cassidy12}. In particular, it has been possible to selectively populate individual Rydberg-Stark states~\cite{Cassidy15} through a two-step excitation scheme $\text{Ps}(1s)\to \text{Ps}(2p)\to \text{Ps}(ns,nd)$~\cite{Ziock90,Cassidy12,Jones14}. These developments make further experimentation feasible, with a view to creating a focused Ps beam suitable for gravity measurements~\cite{Mills02,Cassidy14}. Rydberg Ps is also important for the production of low-energy antihydrogen atoms ($\Hbar$) through collisions with antiprotons ($\overline{p}$)~\cite{Deutch88,Mitroy95}, viz.,
\begin{equation}\label{eq:Ps_p_Hbar}
\text{Ps} + \overline{p} \longrightarrow \Hbar + e^-,
\end{equation}
where the $\Hbar$ production increases rapidly with the excitation of Ps. This reaction is to be used to create antihydrogen in the proposed GBAR~\cite{Sacquin14} and AEgIS~\cite{Doser2012} experiments, designed to test whether the weak equivalence principle applies to antimatter in the same way it does to matter. A number of calculations of reaction~(\ref{eq:Ps_p_Hbar}) have been performed, mostly for ground-state Ps and ground-state $\Hbar$~\cite{Mitroy97}, though there are some calculations for the collisions involving excited states~\cite{Igarashi94,Mitroy95,KAdyrov15,Rawlins2016}.

Recent technological developments~\cite{Danielson15} in positron trapping and detection have made the method we propose here more feasible. The basic procedure is as follows: a time-focused positron pulse is implanted into a suitable material, resulting in the production of ground-state orthopositronium ($o$-Ps) atoms. These are subsequently excited via $n=2$ to levels with $n=3$--30 using nanosecond-pulsed UV ($\lambda = 243.0$~nm) and IR ($\lambda = 729$--1312~nm) laser radiation. Ps atoms in varying Rydberg states and having kinetic energies in the range of 10--1000 meV collide with neutral atoms $A$ in a scattering cell, enabling the reaction
\begin{equation}\label{eq:reac}
\text{Ps}(nl) + A \longrightarrow e^+ A + e^-
\end{equation}
to take place, where $e^+A$ is the positron bound state with the atom, e.g.,
Mg, Cu, or Zn~\footnote{For atoms with positive electron affinities, another
possible charge transfer reaction is $\text{Ps} + A \rightarrow A^- + e^+$.
Formation of the negative $\mbox{Ps}$ ion, with binding energy 0.326~eV
\cite{Drake05}, $\text{Ps} + A \rightarrow \mbox{Ps}^- + A^+$, may also occur,
though likely with very small probability for high-$n$ Rydberg Ps.}. The cross
section for this process depends on the incident Ps energy, the initial
state $nl$ of Ps, and on the positron-atom binding energy $\eb $. Reaction~(\ref{eq:reac}) leads to rapid positron annihilation; the positron-atom
bound state lifetime is~\cite{Gribakin01,Mitroy02a}
\begin{equation}\label{eq:lifetime}
\tau _a\sim 0.7 \eb^{-1/2}~\text{ns,}
\end{equation}
where $\eb $ is in electronvolts. These are typically a few nanoseconds, which is much shorter than Rydberg Ps fluorescence lifetimes. Thus, the formation of bound states in the proposed experiment can be detected by an increase in annihilation events in the scattering cell, and a corresponding decrease in events seen downstream. Varying the Rydberg Ps states and kinetic energies will provide additional controls and make it possible to test theoretical predictions.    

Detection of positron-atom bound states in reaction~(\ref{eq:reac}) will be
the first observation of its kind. A comparison of the measured cross section with the theoretical results derived in this paper should provide an estimate of the positron binding energy, which could be compared with existing high-quality predictions~\cite{Mitroy02,Dzuba12,Harabati14}. It would also be interesting to apply this method to molecules for which the binding energies are known from the resonant annihilation studies
\cite{Gribakin10,Danielson09,Danielson10,Jones12,Danielson12}.
Unlike positron-molecule annihilation which probes resonant, quasibound
states, the molecular analog of reaction~(\ref{eq:reac}) should lead to population of the true positron-molecule bound states. Molecules also allow one to explore reaction~(\ref{eq:reac}) for systems with very small binding energies, e.g.,
$\text{C}_2 \text{H}_6$ or $\text{CH}_3 \text{F}$. Their positron
affinities are expected to be ${\sim} 1$~meV~\cite{Danielson10a,Gribakin06a}
but have not be measured directly because such shifts of the annihilation
resonances are much smaller than the energy resolution of the positron beam.

There are several calculations of the cross section for reaction~(\ref{eq:reac}) and its negative-ion analog. All of them consider the equivalent processes involving the hydrogen or antihydrogen atoms,
\begin{subequations}\label{eq:Hreac}
\begin{gather}
\text{Ps}(nl) + \Hbar \longrightarrow e^+ \Hbar + e^-,\label{eq:Hreac1} \\
\text{Ps}(nl) + \text{H} \longrightarrow \text{H}^- + e^+,\label{eq:Hreac2}
\end{gather}
\end{subequations}
for low $n$. Biswas~\cite{Biswas01} estimated the cross section for Ps($1s$)-H($1s$) collisions using the two-coupled-channel (2CC) formalism, treating the outgoing positron as a plane wave. Later, Blackwood \textit{et al.}~\cite{Blackwood02} and Walters \textit{et al.}~\cite{Walters04} intimated that inclusion of the Coulomb interaction between the ion and lepton in the final state is important for obtaining accurate results. Roy \textit{et al.}~\cite{Roy05} then calculated the cross section for Ps($1s$)-H($1s$) collisions within the Coulomb-modified eikonal approximation (CMEA), which accounts for this Coulomb interaction; they obtained results significantly at variance with those of Biswas~\cite{Biswas01}. Roy and Sinha~\cite{Roy08} extended the work of Roy \textit{et al.}~\cite{Roy05} to include the $n=2$ states of Ps. Most recently, Comini and Hervieux~\cite{Comini13} and Comini \textit{et al.}~\cite{Comini14} computed the cross section for Ps($nl$)-$\Hbar(n'l')$ collisions using the continuum-distorted-wave--final-state (CDW-FS) method; they considered $n=1$--3 and $n'=1$--5. 

Additionally, there exist calculations~\cite{Choudhury86,Straton91,McAlinden02,Ghosh04} for the reverse reactions
\begin{subequations}\label{eq:rev_reac}
\begin{gather}
 e^+ \Hbar + e^- \longrightarrow \text{Ps}(nl) + \Hbar \label{eq:rev_reac1}, \\
 \text{H}^- + e^+ \longrightarrow \text{Ps}(nl) + \text{H} ,\label{eq:rev_reac2}
\end{gather}
\end{subequations}
for $n=1$ and 2, and the total for $n\geq 3$~\cite{McAlinden02}. These can be related to the forward cross sections through the principle of detailed balance~\cite{LandauQM}. We are unaware of any calculations of forward or reverse cross sections for specific $n>3$.

Here we provide an approximate theoretical method for estimating the cross section for reaction~(\ref{eq:reac}) for a generic target atom or molecule $A$. Calculations have first been carried out for reactions~(\ref{eq:Hreac}) and benchmarked against the existing data from the literature to investigate the accuracy of our method. Results are then given for the Rydberg Ps collisions for various $e^+ A $ binding energies.

The paper is organized as follows. Section~\ref{sec:theory} describes the theoretical basis of our calculations; numerical results are then presented in Sec.~\ref{sec:numres}. Section~\ref{sec:exp} outlines the experimental procedures that will be involved. We conclude in Sec.~\ref{sec:conc} with a summary of the work.

\section{\label{sec:theory}Theory}

\subsection{Calculation of the cross section}\label{subsec:calc}

We seek to compute the cross sections for reaction~(\ref{eq:reac}), in which a Ps atom with principal quantum number $n$ and orbital quantum number $l$ collides with a stationary atom $A$ (which is at the origin). The center-of-mass momentum of the incident Ps is $\vec{K}$, and the momentum of the outgoing electron is $\vec{k}$. Unless otherwise stated, atomic units are used.

We work in the first Born approximation, taking the motion of the incident Ps and the outgoing electron as plane waves~\footnote{Note that the Born approximation (without exchange) gives zero cross sections for Ps-atom scattering processes in which the initial and final Ps states have the same parity~\cite{Massey54,McAlinden96}. However, for the asymmetric charge-exchange process considered in the present work, the first Born approximaton gives a nonzero result.}. The positron-atom binding energy is
typically small (a fraction of an electronvolt). The wave function of the weakly bound positron is diffuse and located mostly outside the atom; hence, we describe it using the zero-range-potential model~\cite{Demkov88} (see also Refs.~\cite{Gribakin06,Gribakin06a,Gribakin01}).

The amplitude for the process is given by
\begin{equation}\label{eq:A_m}
A_m(\vec{K}) = \iint e^{-i \vec{k} \cdot \vec{r}_1} \varphi_0^* (\vec{r}_2) V(\vec{r}_2) e^{i\vec{K}\cdot\vec{R}} \psi_{nlm}(\vec{r}) \, d^3\vec{r}_1 \, d^3\vec{r}_2 ,
\end{equation}
where $\vec{r}_1$ ($\vec{r}_2$) is the position of the electron (positron)
with respect to the atom regarded as infinitely massive, $\varphi_0$ is the wave function of the bound positron (with energy $\eo=-\eb$), $V$ is the positron-atom interaction (which serves as the perturbation), $\psi_{nlm}$ is the internal Ps wave function (with $m$ the magnetic quantum number), $\vec{R}=(\vec{r}_1+\vec{r}_2)/2$ is the position of the Ps center of mass, and $\vec{r}=\vec{r}_1-\vec{r}_2$ is the position of the electron in Ps relative to the positron. The cross section $\sigma_m(\vec{K})$ is obtained from
\begin{equation}\label{eq:d_sigma_m}
d\sigma_m = \frac{2\pi}{j} \left\lvert A_m \right\rvert^2 \delta \left( \eo + \frac{k^2}{2} + \frac{1}{4n^2} - \frac{K^2}{4} \right) \, d\rho_f ,
\end{equation}
where $j=K/2$ is the flux density of the incident Ps, $d\rho_f = d^3 \vec{k} /(2\pi)^3$ is the density of final states, and the $\delta $ function ensures
energy conservation~\cite{LandauQM}.

Using spherical polar coordinates $(k,\theta_{\vec{k}},\phi_{\vec{k}})$ in $\vec{k}$ space, we have $d^3\vec{k} = k^2 \, dk \, d\Omega_{\vec{k}} = k \, d(k^2/2) \, d\Omega_{\vec{k}}$, where $d\Omega_{\vec{k}}=\sin\theta_{\vec{k}} \, d\theta_{\vec{k}} \, d\phi_{\vec{k}}$ is the solid angle element. Integrating Eq.~(\ref{eq:d_sigma_m}) over $d(k^2/2)$ we find the differential cross section,
\begin{equation}
\frac{d\sigma_m}{d\Omega_{\vec{k}}} = \frac{k}{2\pi^2 K} \left\lvert A_m \right\rvert^2,
\end{equation}
with the energy conservation law
\begin{equation}\label{eq:en_cons}
k=\sqrt{\frac{K^2}{2} - \frac{1}{2n^2} - 2\eo}.
\end{equation}
The total cross section, averaged over the possible magnetic quantum numbers $m$ of the incident Ps, is then
\begin{equation}
\sigma = \frac{1}{2l+1} \frac{k}{2\pi^2 K} \sum_{m=-l}^l \int \left\lvert A_m \right\rvert^2 \, d\Omega_{\vec{k}}.
\end{equation}

To determine the amplitude $A_m$, Eq.~(\ref{eq:A_m}), we use the Schr\"odinger
equation for the bound positron, $\varphi_0^*(\vec{r}_2)V(\vec{r}_2) = \bigl(\frac{1}{2} \nabla_2^2 + \eo\bigr)\varphi_0^*(\vec{r}_2)$, where the wave function behaves as $\varphi_0(\vec{r}_2) \simeq B e^{-\kappa r_2}/r_2$ at large $r_2$, $\kappa=\sqrt{-2\eo}$, and $B$ is a normalization constant. It is convenient to express the internal Ps wave function $\psi_{nlm}$ in terms of its momentum-space counterpart $\widetilde{\psi}_{nlm}$, viz.,
\begin{equation}
\psi_{nlm}(\vec{r}) = \int e^{i\vec{q}\cdot\vec{r}} \widetilde{\psi}_{nlm}(\vec{q}) \frac{d^3 \vec{q}}{(2\pi)^3},
\end{equation}
so that
\begin{align*}
A_m &= \int \frac{d^3\vec{q}}{(2\pi)^3} \widetilde{\psi}_{nlm} (\vec{q})
\int d^3\vec{r}_1  \exp\left[ i \left( -\vec{k} + \frac{\vec{K}}{2} + \vec{q}\right) \cdot\vec{r}_1 \right]
\nonumber\\
&\times \int d^3\vec{r}_2 \exp \left[ i \left( \frac{\vec{K}}{2} - \vec{q} \right)\cdot\vec{r}_2 \right] \left( \frac12 \nabla^2_2 - \frac{\kappa^2}{2} \right)\varphi_0^*(\vec{r}_2).
\end{align*}
The integral over $\vec{r}_1$ yields $(2\pi)^3 \delta(-\vec{k}+\vec{K}/2+\vec{q})$. Invoking the Hermiticity of the Laplacian operator gives
\begin{align*}
A_m &= -\frac12 \widetilde{\psi}_{nlm}\left( \vec{k}-\frac{\vec{K}}{2} \right) \nonumber\\
&\times \int \varphi_0^*(\vec{r}_2) \left( \kappa^2 + \left\lvert \vec{K}-\vec{k} \right\rvert^2 \right) \exp[i(\vec{K}-\vec{k})\cdot\vec{r}_2] \, d^3\vec{r}_2.
\end{align*}
Defining $\widetilde{\varphi}_0(\vec{q}) \equiv \int e^{-i\vec{q}\cdot\vec{r}} \varphi_0(\vec{r}) \, d^3\vec{r}$, and adopting the zero-range-model approximation in which $\varphi_0(\vec{r}_2) = B e^{-\kappa r_2}/r_2$ in all space,
we have
\begin{equation}
\widetilde{\varphi}_0(\vec{q}) = B \int \frac{e^{-\kappa r}}{r} e^{-i\vec{q}\cdot\vec{r}} \, d^3\vec{r}
= \frac{4\pi B}{\kappa^2 + q^2},
\end{equation}
and $B=\sqrt{\kappa/2\pi}$. Thus, we finally obtain
\begin{align}
A_m = -\sqrt{2 \pi\kappa}  \widetilde{\psi}_{nlm}  \left( \vec{k}-\frac{\vec{K}}{2} \right),
\end{align}
which gives
\begin{equation}\label{eq:cs}
\sigma = \frac{1}{2l+1} \frac{k\kappa}{\pi K} \sum_{m=-l}^l \int \left\lvert \widetilde{\psi}_{nlm}  \left( \vec{k}-\frac{\vec{K}}{2} \right) \right\rvert^2\, d\Omega_{\vec{k}}.
\end{equation}

The internal Ps wave function in momentum space, $\widetilde{\psi}_{nlm}$, can be written as
\begin{equation}\label{eqn:FYprod}
\widetilde{\psi}_{nlm}(\vec{p}) = (2\pi)^{3/2} F_{nl}(p) Y_{lm}(\hat{\vec{p}}),
\end{equation}
where $Y_{lm}$ is a spherical harmonic, and
\begin{align}\label{eq:Fnl}
F_{nl}(p) &= \left( \frac12 \right)^{-3/2} \left[ \frac{2}{\pi} \frac{(n-l-1)!}{(n+l)!} \right]^{1/2} n^2 2^{2l+2} l! \nonumber\\
&\times \frac{(2np)^l}{\left[ (2np)^2 + 1 \right]^{l+2}} C_{n-l-1}^{(l+1)} \left( \frac{(2np)^2-1}{(2np)^2+1} \right),
\end{align}
with $C_\nu^{(\alpha)}$ being a Gegenbauer polynomial~\cite{BraJoa}. Substituting
Eq.~(\ref{eqn:FYprod}) into Eq.~(\ref{eq:cs}) and invoking the addition theorem for spherical harmonics gives
\begin{equation}\label{eq:cs2}
\sigma = \frac{2\pi k\kappa}{K} \int \left\lvert F_{nl}\left( \left\lvert \vec{k}-\frac{\vec{K}}{2} \right\rvert \right) \right\rvert^2 d\Omega_{\vec{k}}.
\end{equation}
Choosing the incident Ps momentum $\vec{K}$ along the $z$ axis means that the integrand in Eq.~\eqref{eq:cs2} has no dependence on the azimuthal angle $\phi_{\vec{k}}$. Therefore,
\begin{equation*}
\sigma=\frac{4\pi^2 k\kappa}{ K} \int_0^\pi \left\lvert F_{nl}\left( \sqrt{k^2+\frac{K^2}{4}-kK\cos\theta_{\vec{k}}} \right) \right\rvert^2 \sin\theta_{\vec{k}} \, d\theta_{\vec{k}},
\end{equation*}
and making the substitution $p=(k^2+K^2/4-kK\cos\theta_{\vec{k}})^{1/2}$, we
find
\begin{equation}\label{eq:cs3}
\sigma=\frac{8\pi^2\kappa}{ K^2} \int_{\left\lvert k-K/2\right\rvert}^{k+K/2} p \left\lvert F_{nl}(p) \right\rvert^2 \, dp.
\end{equation}

Note that the cross section is proportional to the probability of finding the
electron with momentum $\vec{p}=\vec{k}-\vec{K}/2$ in the initial Ps state
[see Eq.~(\ref{eq:cs}) or (\ref{eq:cs2})]. 
This is the momentum that must be added to the average momentum of the
electron within the incident Ps ($\vec{K}/2$) to create an outgoing electron
with momentum $\vec{k}$, i.e., $\vec{p}$ is the momentum transfer.

Before looking at numerical values of the cross section~(\ref{eq:cs3}), there is an important point to note concerning the energy conservation relation~(\ref{eq:en_cons}). For $\eb > 1/4n^2$, the reaction is exothermic and feasible for any Ps momentum $K$. Conversely, for $\eb < 1/4n^2$ the reaction is endothermic and only feasible for $K>K_\text{th}$, where $K_\text{th}$ is the threshold Ps
momentum,
\begin{equation}
K_\text{th} = \sqrt{\frac{1}{n^2} - 4\eb}.
\end{equation}
For a fixed binding energy $\eb$ there is a critical value
$n_{\text{crit}}=(4\eb )^{-1/2}$ such that the reaction is endothermic for the Ps principal quantum numbers $n<n_\text{crit}$ and exothermic for $n>n_\text{crit}$.

In the exothermic case the cross section~(\ref{eq:cs3}) behaves as
$\sigma \propto 1/K$ near threshold ($K\rightarrow 0$), while in the endothermic
case one has $\sigma \propto k\propto \sqrt{E-E_\text{th}}$, when the Ps center-of-mass-motion energy, $E=K^2/4$, is close to the threshold energy $E_\text{th}=1/4n^2-\eb $. Such behaviour is in agreement with the Wigner threshold laws for particles with
short-range interactions~\cite{LandauQM}.
In a more accurate treatment, the Coulomb interaction between the final-state
electron and $e^+A$ must be included, which would change the latter threshold law to $\sigma =\text{const}$.
As we will see in Sec.~\ref{sec:numres}, in the $\eb < 1/4n^2$ case, a rapid
rise of the cross section from threshold quickly turns into decrease. One can
thus expect that the effect of the Coulomb interaction is small outside the
narrow near-threshold region in which the electron's kinetic energy is smaller than the
Coulomb interaction in the initial Ps state, i.e., for $k^2/2<1/r\sim 1/2n^2$
(using the mean Coulomb interaction in Ps($nl$) in the last estimate).

\subsection{Semiclassical approximation}\label{subsec:sc}

Although it is straightforward to calculate the cross section by evaluating the integral in Eq.~(\ref{eq:cs3}) numerically (see Sec.~\ref{sec:numres}), an approximate analytical solution can be derived by invoking a semiclassical approximation. This leads to a simple expression for the cross section and
provides additional physical insight into the nature of the problem.

The quantity $p^2 \lvert F_{nl}(p) \rvert^2$ is the probability density of the internal momentum of the incident Ps. For large principal quantum numbers $n$, the motion in the Coulomb field  can be described semiclassically~\cite{LandauQM}. The Ps Rydberg states produced by two-photon excitations~\cite{Cassidy12,Cassidy15} have $l=0$, 2. The simplest answer for $n\gg l$ can be obtained by replacing $p^2 \lvert F_{nl}(p) \rvert^2$ by its classical counterpart $w_n(p)$ for zero classical angular momentum ($L=0$) (see, e.g., Ref.~\cite{Korsch00}):
\begin{equation}\label{eq:w_n}
w_n(p) = \frac{4p_n^3}{\pi\left(p^2+p_n^2\right)^2},
\end{equation}
where $p_n=\sqrt{-2\mu E_n}$, with $\mu=\frac12$ the reduced mass of Ps and $E_n=-1/4n^2$ the quantized Ps energy levels.

Note that the classical angular momentum $L$ is related to the orbital quantum number $l$ by $L=l+\frac12$. In principle, one could calculate the cross sections for $l=0$ and 2 using the generic classical momentum distribution for $L\geq 0$ (see Ref.~\cite{Korsch00}) instead of Eq.~(\ref{eq:w_n}). In this case, however, the cross section does not have a simple analytical form. As we will see in Sec.~\ref{sec:numres}, the semiclassical cross section derived from Eq.~(\ref{eq:w_n}) is a good approximation for low $l$, such as $l=0$ and 2.

Substituting $w_n(p)/p^2$ in place of $\lvert F_{nl}(p) \rvert^2$ in Eq.~(\ref{eq:cs3}) we have
\begin{equation}\label{eq:sig_sc0}
\sigma=\frac{32\pi \kappa p_n^3}{K^2} \int_{\lvert k-K/2\rvert}^{k+K/2} \frac{dp}{p\left(p^2+p_n^2\right)^2},
\end{equation}
which gives the semiclassical cross section
\begin{align}
\sigma=\frac{16\pi\kappa}{p_n K^2} &\left[ \ln\left(\frac{k+K/2}{k-K/2}\right)^2+\ln \frac {(k-K/2)^2+p_n^2}{(k+K/2)^2+p_n^2}\right. \notag  \\
&\left. {}+\frac{p_n^2}{(k+K/2)^2+p_n^2}-
\frac{p_n^2}{(k-K/2)^2+p_n^2} \right].\label{eq:sig_sc}
\end{align}
Note that this expression diverges weakly (logarithmically) for $k=K/2$. This
occurs for endothermic reactions at the incident Ps momentum
$K=\sqrt{2}K_\text{th}$. The corresponding peak in the semiclassical cross
section coincides with the maximum of the $l=0$ quantum-mechanical cross section
(see Sec.~\ref{sec:numres}).

\section{\label{sec:numres}Numerical results}

Cross sections have been computed in the present work by evaluating Eq.~(\ref{eq:cs3}) numerically. The functions $F_{nl}(p)$ were computed for $0\leq p\leq15$ on a linear grid with 1,000,000 points. For each combination of $n$ and $l$, the accuracy of the procedure was tested by evaluating the normalization integral
\begin{equation} \label{eq:normalization}
I=\int_0^\infty \lvert F_{nl}(p) \rvert^2 p^2 \, dp
\end{equation}
numerically. In every case, the computed value of $I$ was found to be within $10^{-8}$ of the exact value $I=1$. With the linear grid it is necessary to use such a large number of points because the function $F_{nl}$ possesses $n-l-1$ nodes, so for Rydberg states with large $n$ and small $l$, $F_{nl}(p)$ oscillates rapidly at small $p$.

\subsection{Comparisons with existing calculations for Ps-H collisions}

The aim of the theoretical part of this work is to obtain estimates of the cross sections of reaction (\ref{eq:reac}) for high Rydberg states of Ps and weakly-bound positron states ($\eb <0.5$~eV), for which the approximations used in Sec.~\ref{subsec:calc} are justified. Since there are no previous calculations of this process, the only comparison that can be made is with a number of calculations for reactions (\ref{eq:Hreac}) involving (anti)hydrogen and incident Ps with $n=1$--3. When examining these results, one should have in mind that our method is by far the simplest, and that it is not expected to be accurate for low $n$ and the relatively strongly bound $\text{H}^-$ or $e^+\Hbar$ ($\eb =0.754$~eV~\cite{Andersen99}). What we are looking for here is a broad order-of-magnitude agreement and correct energy dependence of the cross sections (except in the narrow near-threshold energy range).

To account for the fact that H$^-$ can only form in reaction (\ref{eq:Hreac2}) if the total electron spin is zero, the cross sections~(\ref{eq:cs3}) are multiplied by a factor of $\frac{1}{4}$. Also, using $\eb =0.0277$~a.u. gives $B=\sqrt{\kappa /2\pi }\approx 0.1936$. However, the true value of $B$, extracted from the asymptotic form of the accurate wave function is 0.3159~\cite{Frolov04}. Therefore, we have multiplied the cross sections~(\ref{eq:cs3}) by a extra factor of $(0.3159/0.1936)^2 \approx 2.66$.

\begin{figure}
\includegraphics*{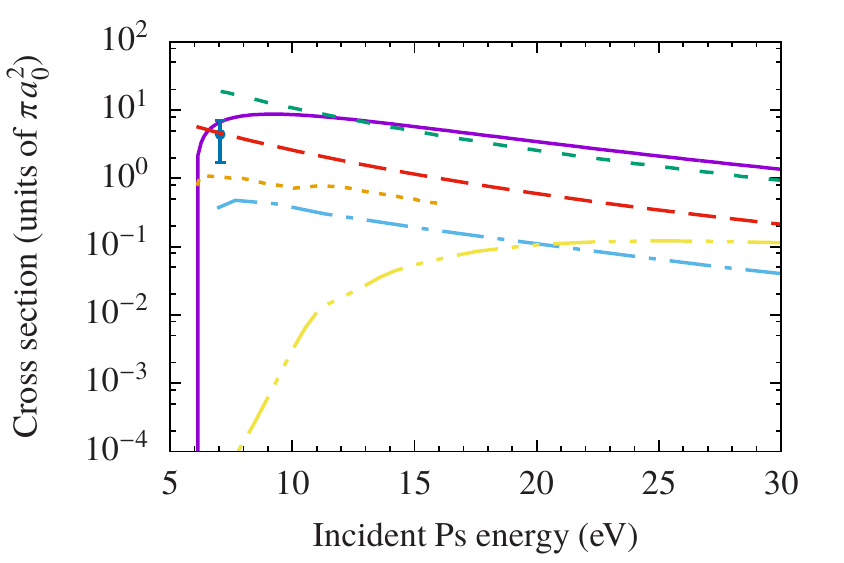}
\caption{\label{fig:1s_comp}Cross sections for H$^-$ formation in Ps($1s$)-H collisions. Solid purple curve, present; blue circle with error bars, CBA~\cite{Straton91}; dot-dash-dotted yellow curve, 2CC~\cite{Biswas01}; dot-dashed blue curve, CMEA~\cite{Roy05}; short-dashed green curve, CDW-FS~\cite{Comini14}; dotted orange curve, CPA~\cite{McAlinden02}; long-dashed red curve, DWBA (see text).}
\end{figure}

Figure~\ref{fig:1s_comp} compares our calculation for Ps($1s$) with the existing calculations.
The computations by McAlinden \textit{et al.}~\cite{McAlinden02} were for the reverse reaction~(\ref{eq:rev_reac2}), which we have converted into forward cross sections through the principle of detailed balance~\cite{LandauQM}:
\begin{equation}\label{eq:db}
\sigma_{(\text{\ref{eq:Hreac2}})} = \frac{k^2}{4K^2(2l+1)}\sigma_{(\text{\ref{eq:rev_reac2}})} .
\end{equation}
These reverse cross sections were obtained from coupled-pseudostate-approach (CPA) calculations in which one of the electrons was kept `frozen' in the $1s$ state of the H atom and only the Ps pseudostates were included. The reaction for Ps($1s$) is endothermic, with a threshold incident Ps energy of 6.05~eV. The present results and the 2CC results of Biswas~\cite{Biswas01} do not include the Coulomb interaction between the ion and outgoing positron; hence they show zero cross section at threshold energy. On the other hand, the CPA results of McAlinden \textit{et al.}~\cite{McAlinden02}, CMEA calculations by Roy \textit{et al.}~\cite{Roy05}, and the CDW-FS method of Comini and Hervieux~\cite{Comini13} do account for this Coulomb interaction, leading to finite cross sections at threshold. 
Also shown are the earlier Coulomb-Born approximation (CBA) results of Straton and Drachman~\cite{Straton91}, who obtained a range of cross section values, using various $\text{H}^-$ states and orthogonalization corrections at selected energies. (We have ignored one of their values that was an order or magnitude above the rest.) In addition, we calculated the Ps formation cross section from $\text{H}^-$ using distorted-wave Born approximation (DWBA), which gives the Ps formation cross section in He with $20\%$ accuracy, though overestimates it for heavier noble gases by a larger amount~\cite{DunlopPhD,Dunlop06}. The corresponding cross section obtained from Eq.~(\ref{eq:db}) is also shown in Fig.~\ref{fig:1s_comp}.

As seen in Fig.~\ref{fig:1s_comp}, there are significant discrepancies between the various calculations. The energy dependence of the 2CC result of Biswas~\cite{Biswas01} makes it an outlier. Other calculations show similar energy dependence, though absolute values differ by an order of magnitude. It is in fact remarkable that the results of present approach fit within the range of values from other, more sophisticated methods, in spite of the fact that it is not expected to work for $\text{Ps}(1s)$.

\begin{figure}
\includegraphics*{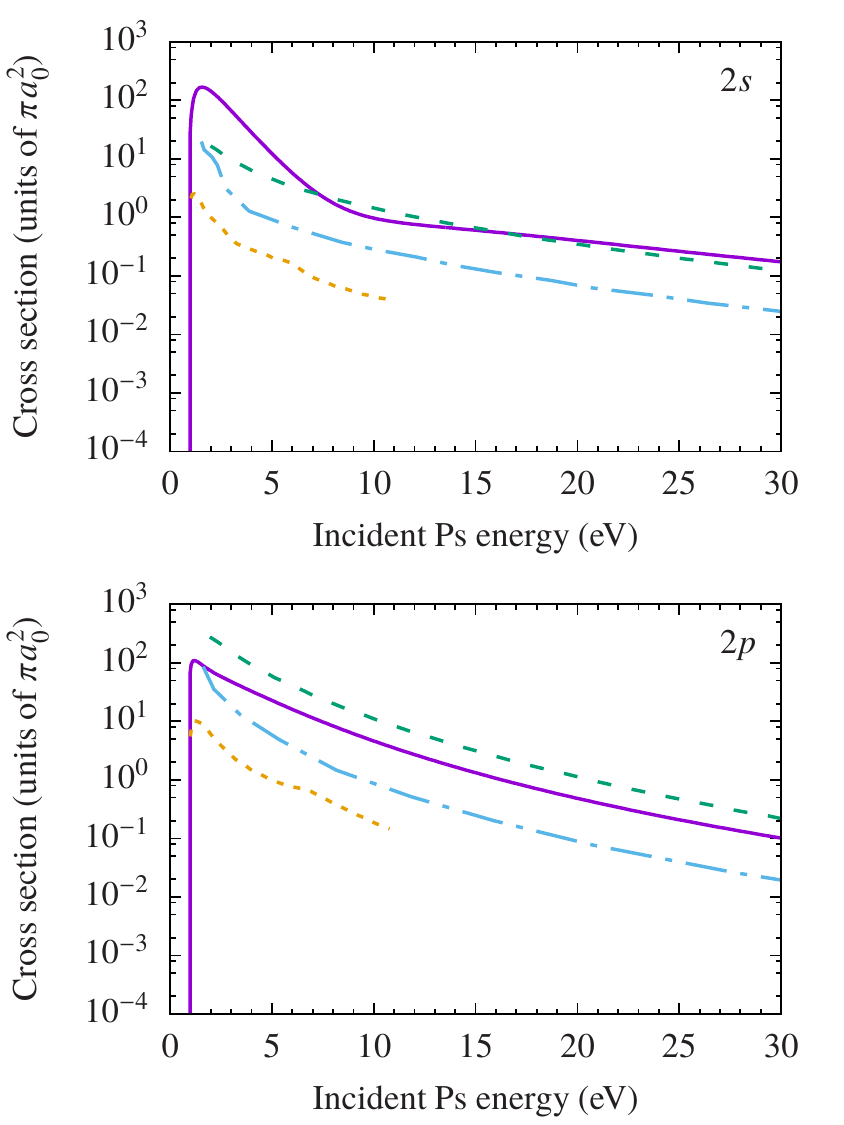}
\caption{\label{fig:2s2p_comp}Cross sections for H$^-$ formation in Ps($2s$,$2p$)-H collisions. Solid purple curves, present; dot-dashed blue curves, CMEA~\cite{Roy08}; short-dashed green curves, CDW-FS~\cite{Comini13}; dotted orange curves, CPA~\cite{McAlinden02}.}
\end{figure}

\begin{figure}
\includegraphics*{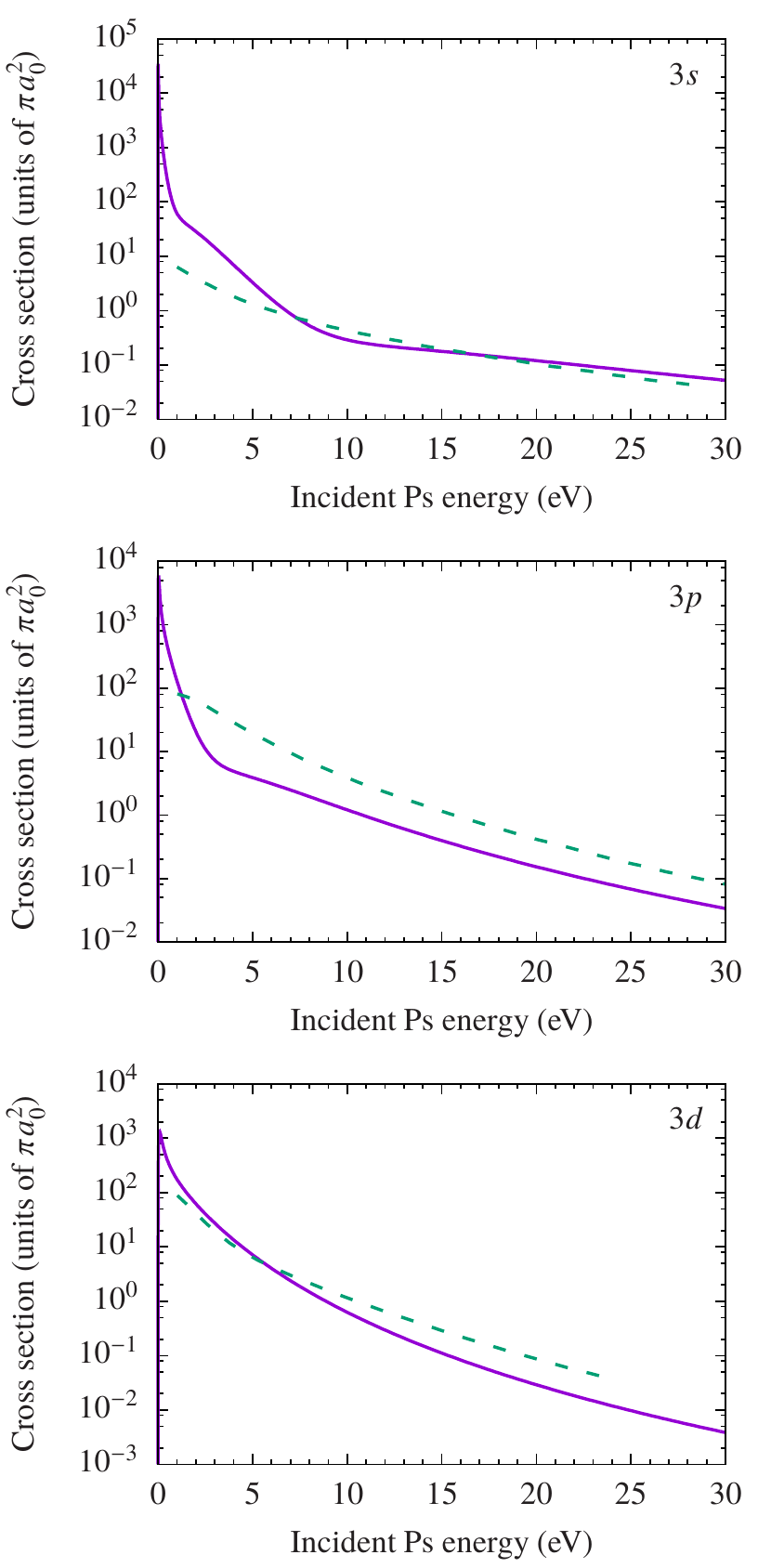}
\caption{\label{fig:3s3p3d_comp}Cross sections for H$^-$ formation in Ps($3s$,$3p$,$3d$)-H collisions. Solid purple curves, present; short-dashed green curves, CDW-FS~\cite{Comini13}.}
\end{figure}

Figure~\ref{fig:2s2p_comp} compares the present calculations for Ps($2s$,$2p$) with the existing calculations.
Here the CPA cross sections of McAlinden~\textit{et al.}~\cite{McAlinden02} are lowest in magnitude, while the other three methods are in a better overall agreement. The present results are in fact quite close to the CDW-FS calculations~\cite{Comini13}. A similar level of agreement can also be seen in Fig.~\ref{fig:3s3p3d_comp}, which compares our cross sections for Ps($3s$,$3p$,$3d$) with the only other available set of results by Comini and Hervieux~\cite{Comini13}.

The level of agreement with existing calculations observed in Figs.~\ref{fig:1s_comp}--\ref{fig:3s3p3d_comp}, especially for Ps($n=3$) states, confirms that our approach should be suitable for making estimates of the charge exchange cross sections of processes involving Rydberg Ps.

\subsection{Predictions for the formation of positron bound states}\label{subsec:pred}

We now present our cross sections for the formation of positron bound states in collisions between Rydberg Ps and atoms (or molecules) that would be used in the experiment. As explained in Sec.~\ref{sec:intro}, the two-step excitation scheme produces Ps in $s$ and $d$ states, and we carry out calculations for $l=0$, $n=1$--20, and $l=2$, $n=3$--20. The only parameter that characterizes the positron bound state is its binding energy. We use the following values: 0.464~eV for Mg~\cite{Bromley06}, 0.170~eV for Cu~\cite{Dzuba99}, 0.107~eV for Zn~\cite{Harabati14}, 0.01~eV for $\text{C}_2 \text{H}_6$~\cite{Barnes03}, and $3\times 10^{-4}$~eV for $\text{CH}_3 \text{F}$~\cite{Gribakin06a}. For a more complete picture, we also consider a species with the binding energy of 0.04~eV, e.g., as measured for $\text{CH}_3 \text{Br}$~\cite{Danielson10a}. 

\begin{figure*}
\includegraphics*{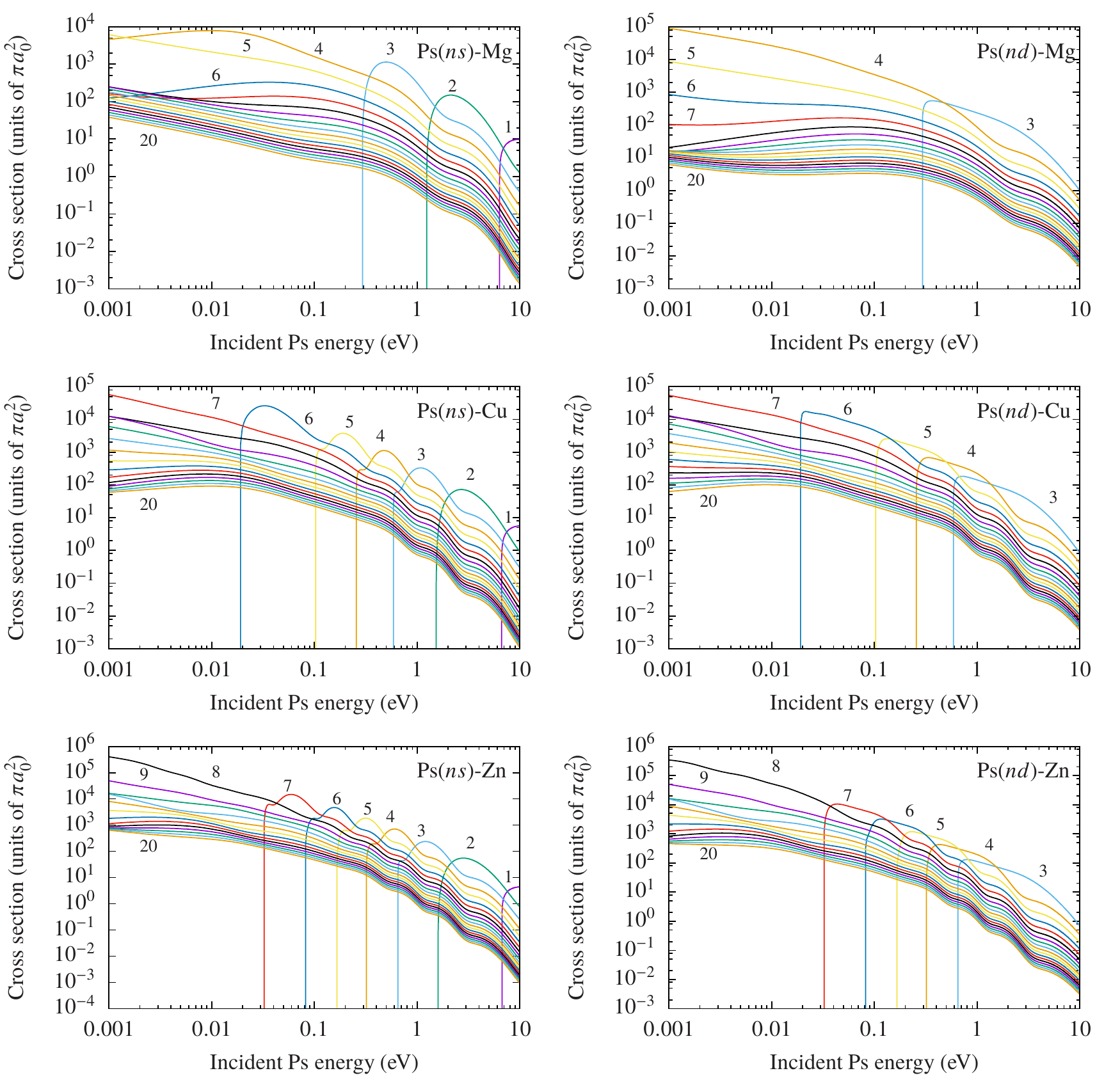}
\caption{\label{fig:CS_MgCuZn}Cross sections for the formation of positron-atom bound states in Ps($ns$,$nd$) collisions with Cu, Mg, and Zn. To identify the value of $n$ to which each curve corresponds, consider the incident Ps energy of 10~eV. At this energy, the cross section decreases monotonically with increasing $n$. Selected values of $n$ are shown explicitly next to the corresponding curves.}
\end{figure*}

\begin{figure*}
\includegraphics*{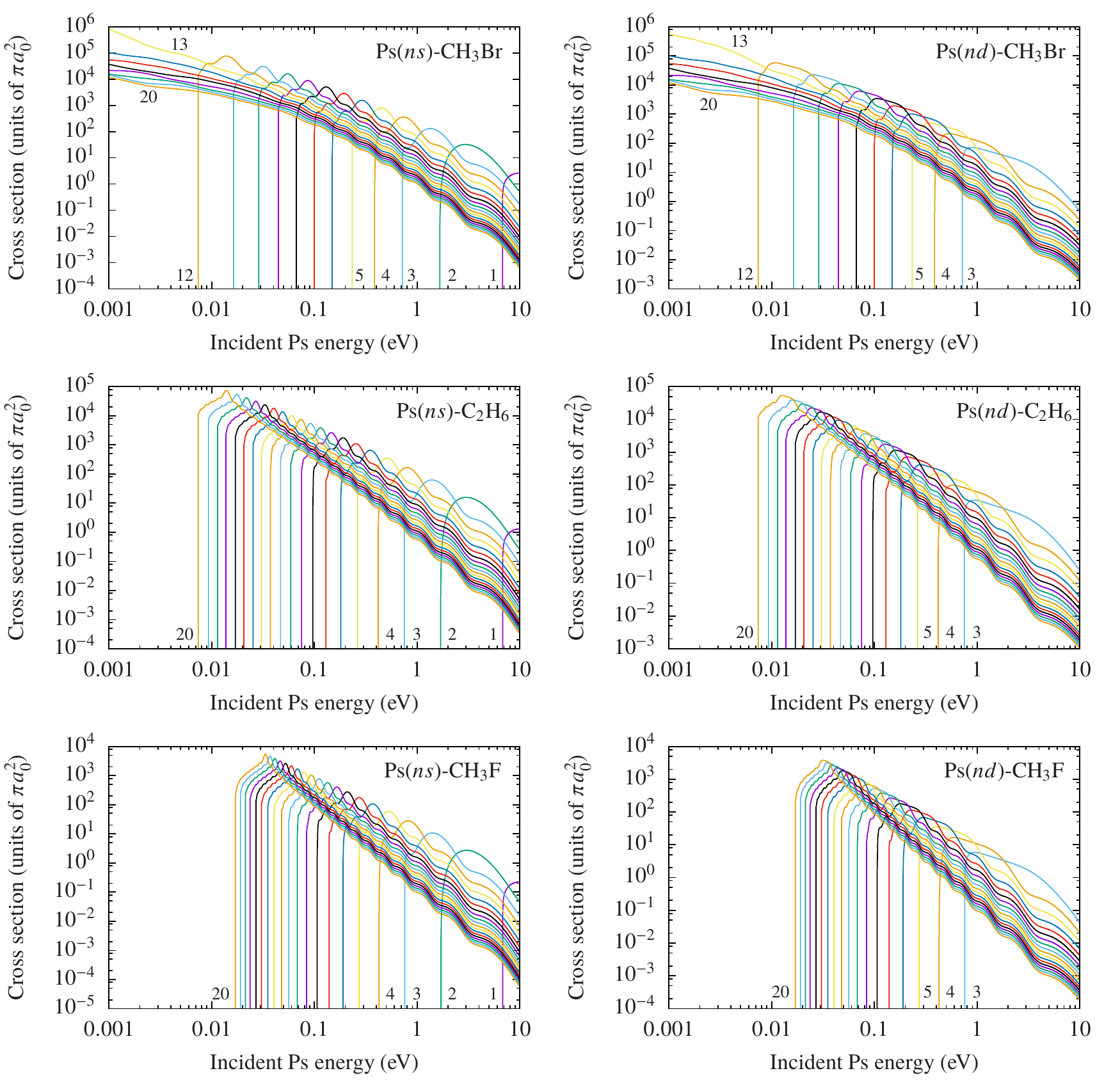}
\caption{\label{fig:CS_molecules}Cross sections for the formation of positron-molecule bound states in Ps($ns$,$nd$) collisions with $\text{C}_2 \text{H}_6$, $\text{CH}_3 \text{F}$, and $\text{CH}_3 \text{Br}$. To identify the value of $n$ to which each curve corresponds, consider the incident Ps energy of 10~eV. At this energy, the cross section decreases monotonically with increasing $n$. Selected values of $n$ are shown explicitly next to the corresponding curves.}
\end{figure*}

Figures~\ref{fig:CS_MgCuZn} and \ref{fig:CS_molecules} show the results for the atoms and molecules respectively. The cross sections are rather featureless, rising rapidly from threshold, in the endothermic case, and decreasing monotonically past the
maximum. The latter occurs at the Ps energy $E\approx 2E_\text{th}$ for
the Ps($ns$) states (see below), and even closer to threshold for the $nd$
states. In the exothermic case the cross sections typically decrease from
threshold.
In general, the largest cross section in the incident Ps energy range studied (0.001--10~eV) is for $n\approx n_\text{crit}$, i.e., the value of $n$ for which the positron transfer is resonant, so that $1/4n^2\approx \eb $. For $n\geq 3$ for the $ns$ states, and $n\geq 4$ for the $nd$ states, one can also see some oscillations superimposed on the decreasing cross section background. These are caused by an oscillatory behaviour of the integrand in Eq.~(\ref{eq:cs3}) and the positions of its maxima and minima in relation to the integration limits.

\begin{figure*}
\includegraphics*{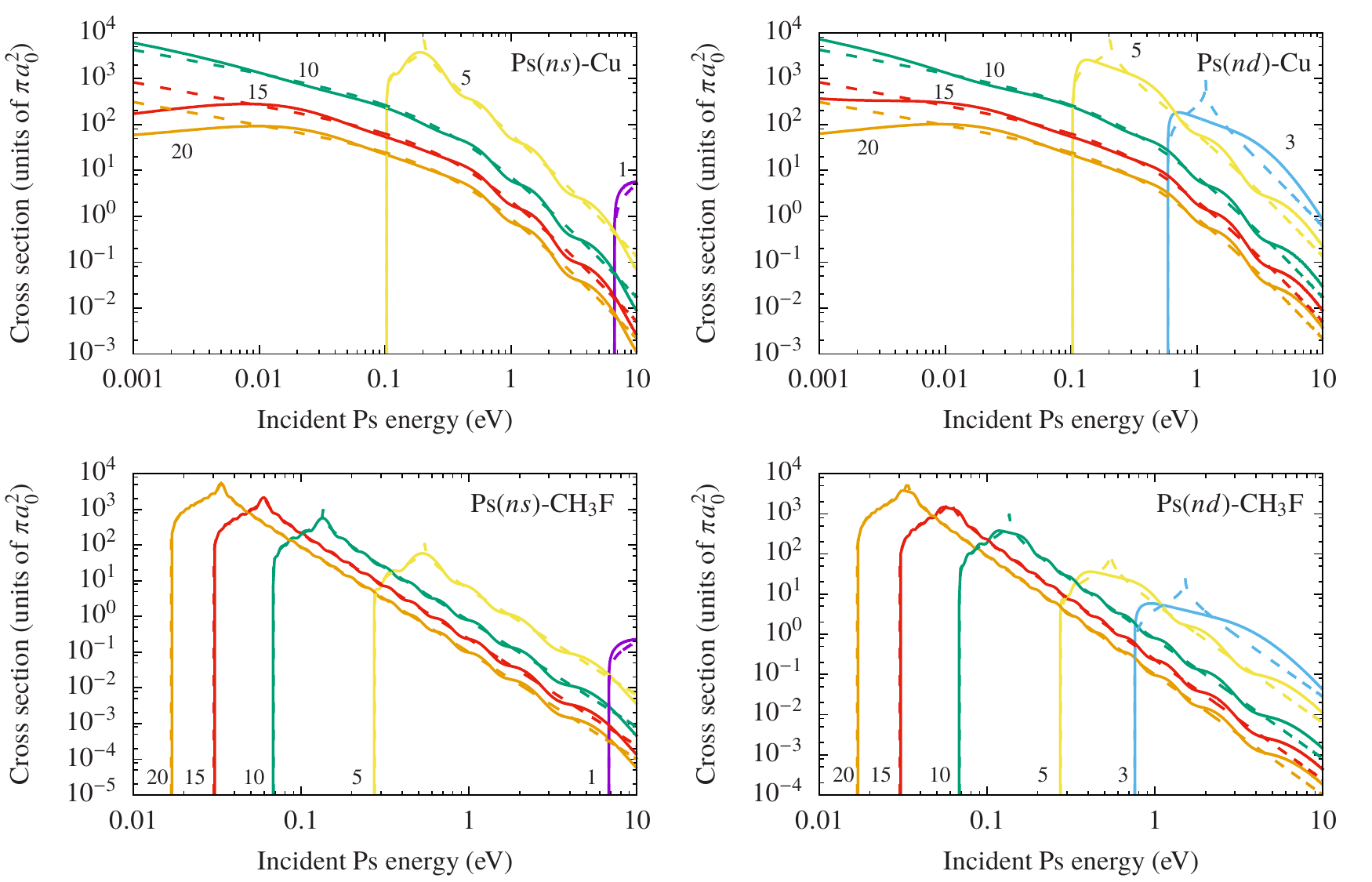}
\caption{\label{fig:cs_semiclassical}Cross sections for positron-bound-state formation in Ps($ns$,$nd$) collisions with Cu and CH$_3$F. Solid curves, quantum-mechanical cross sections, Eq.~(\ref{eq:cs3}); dashed curves, semiclassical cross sections, Eq.~(\ref{eq:sig_sc}). The values of $n$ are shown explicitly.}
\end{figure*}

In Figure~\ref{fig:cs_semiclassical}, we compare the quantum-mechanical cross sections~(\ref{eq:cs3}) for Cu and CH$_3$F with the corresponding semiclassical cross sections~(\ref{eq:sig_sc}) for several values of $n$. 
It can be seen that the agreement is very close for incident Ps energies ${\gtrsim}0.01$~eV, even for low $n$, and particularly for $s$ states of Ps. This comparison also shows that for $l\ll n$ the charge-exchange cross section is almost $l$ independent. The weak singularity of the semiclassical cross sections for endothermic reactions at $E=2E_\text{th}$ coincides with the maximum of the quantum-mechanical cross sections. In both instances this feature is related to the dominant contribution of small momenta $p$ in the case when the lower integration limit in Eqs.~(\ref{eq:cs3}) and (\ref{eq:sig_sc0}) is zero. As expected, 
the semiclassical cross sections obtained by using the monotonic classical momentum distribution~(\ref{eq:w_n}) do not have the oscillatory pattern of their quantum-mechanical counterparts.

\begin{figure}
\includegraphics*{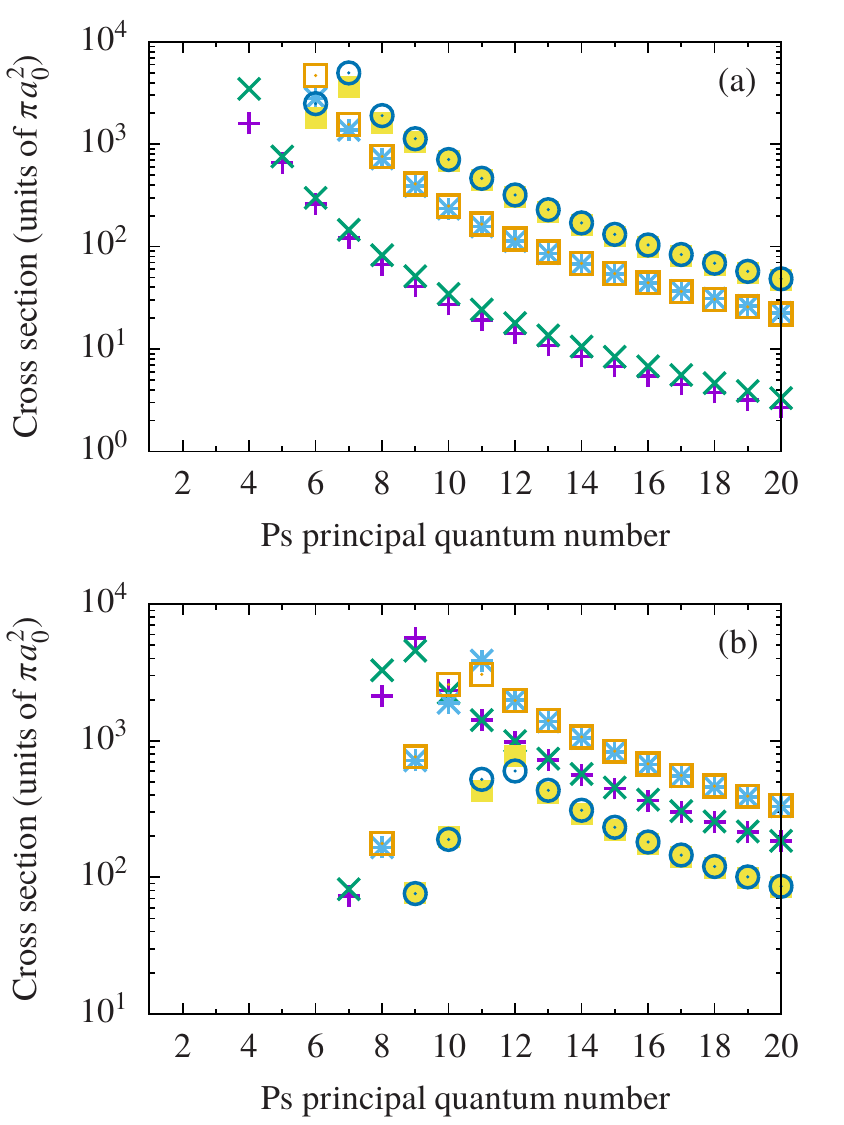}
\caption{\label{fig:n_dep}Dependence of the cross section on $n$ for atoms (a) and molecules (b), at a fixed incident Ps energy of 0.1~eV. In (a): purple plusses, Ps($ns$)-Mg; green crosses, Ps($nd$)-Mg; blue asterisks, Ps($ns$)-Cu; open orange squares, Ps($nd$)-Cu; filled yellow squares, Ps($ns$)-Zn; open blue circles, Ps($nd$)-Zn. In (b): purple plusses, Ps($ns$)-$\text{CH}_3\text{Br}$; green crosses, Ps($nd$)-$\text{CH}_3\text{Br}$; blue asterisks, Ps($ns$)-$\text{C}_2\text{H}_6$; open orange squares, Ps($nd$)-$\text{C}_2\text{H}_6$; filled yellow squares, Ps($ns$)-$\text{CH}_3\text{F}$; open blue circles, Ps($nd$)-$\text{CH}_3\text{F}$.}
\end{figure}

Figure~\ref{fig:n_dep} shows the dependence of the cross sections on the principal quantum number $n$ of the incident Ps for the various atoms and molecules, at a fixed incident Ps energy of 0.1~eV. For systems with larger positron binding energies this dependence is monotonically decreasing. However, when the binding energy drops below 0.1~eV [Fig.~\ref{fig:n_dep}(b)], the $n$ dependence develops a clear maximum.
As the principal quantum number increases, the size of the Ps($nl$) state,
$r\sim 3n^2$, becomes large compared to the size of the positron bound state,
$r_2\sim 1/\kappa =1/\sqrt{2\eb }$. The corresponding $n$ is in fact not so
large, e.g., even for the most weakly bound species
($\text{CH}_3\text{F}$, $\eb =0.3$~meV) this occurs for $n>10$. This means that for large $n$ (and sufficiently large incident Ps energies), the charge-transfer process probes the internal Ps motion at small distances.
Since the Ps wave function depends on $n$ as $\psi_{nlm}(\vec{r})\propto n^{-3/2}$ at small $r$~\cite{LandauQM}, its Fourier transform depends on $n$ as $\widetilde{\psi}_{nlm}(\vec{p})\propto n^{-3/2}$ at large $p$. Consequently, the cross section~(\ref{eq:cs3}) decreases as $\sigma\sim n^{-3}$ at large $n$, as seen
in Fig.~\ref{fig:n_dep}.

\begin{figure}
\includegraphics*{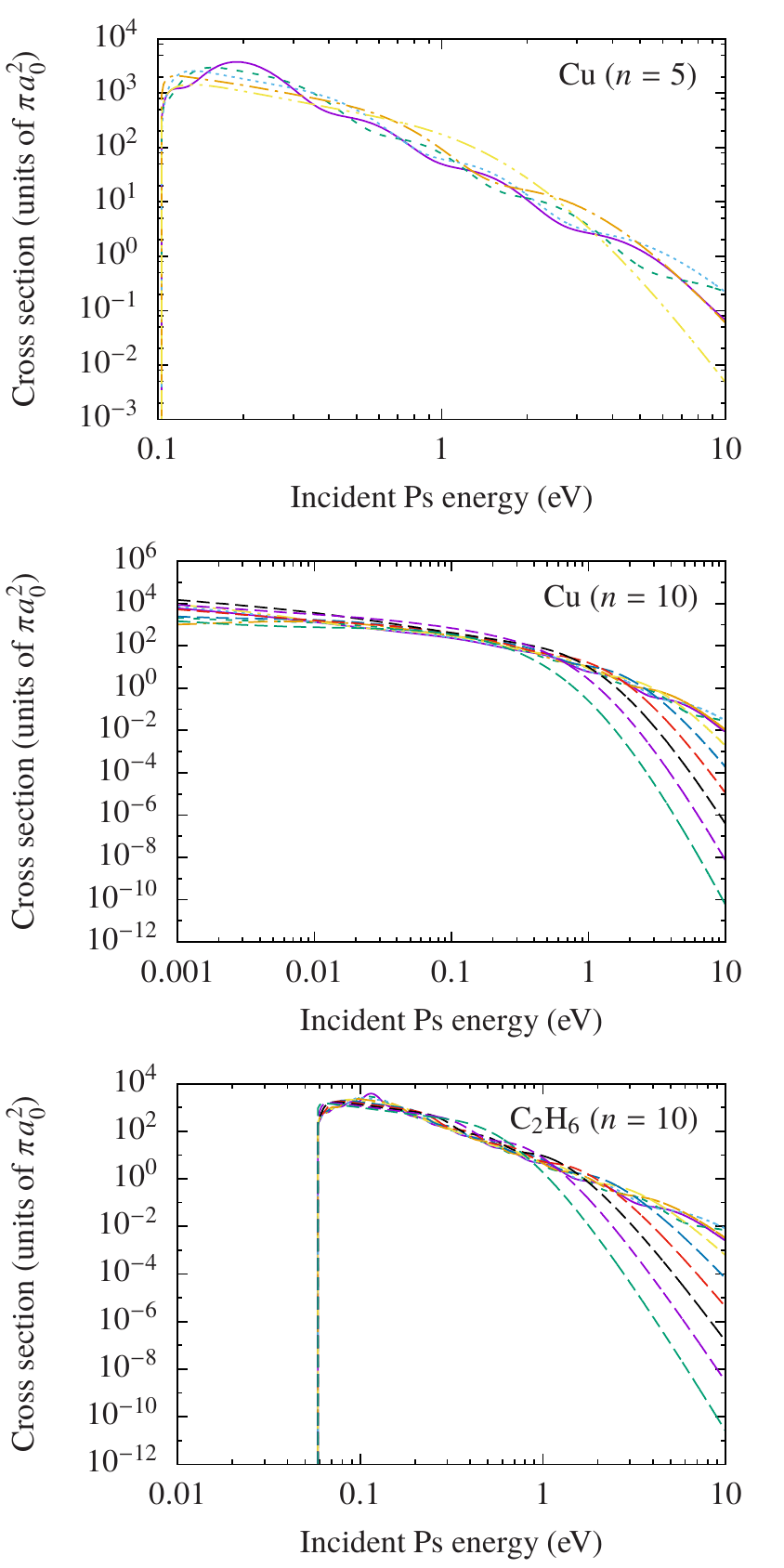}
\caption{\label{fig:l_dep}Cross sections for Ps($nl$) collisions with Cu (with $n=5$ and 10) and $\text{C}_2\text{H}_6$ (with $n=10$). Solid purple curves, $l=0$; short-dashed green curves, $l=1$; dotted light blue curves, $l=2$; dot-dashed orange curves, $l=3$; dot-dash-dotted yellow curves, $l=4$. The long-dashed curves are as follows: dark blue curves, $l=5$; red curves, $l=6$; black curves, $l=7$; purple curves, $l=8$; green curves, $l=9$. For the incident Ps energy of 10~eV and $n=10$, the smallest cross section corresponds to $l=9$, the second smallest to $l=8$, etc., up to $l=4$, below which the cross sections have similar magnitudes.}
\end{figure}

In the experimental setup, the presence of electric fields means that the Ps atom may not be in a pure $s$ or $d$ state, but in a Stark state, i.e., a superposition of states with different $l$. To estimate the importance of this effect we investigate how much the cross sections depend on the orbital quantum number $l$ of the incident Ps state $nl$. Figure~\ref{fig:l_dep} shows the cross sections for Cu and $\text{C}_2\text{H}_6$, for fixed $n$ and $l=0,\dots ,\,n-1$. It can be seen that at low incident Ps energies (e.g., ${{\lesssim} 1}$~eV for $n=10$) the curves for the various values of $l$ are all within an order of magnitude of each other. This indicates that the effect of Stark mixing on the cross sections at low energies is relatively unimportant, i.e., the cross sections for the Stark states and pure $nl$ states should agree to within an order of magnitude.

\FloatBarrier

\section{\label{sec:exp}Experimental procedures}

The experimental production of positron-atom bound states in the way we propose requires a beam of Rydberg Ps atoms that is able to interact with the neutral target atoms in a controlled manner, such that only the bound-state formation causes an increase in the annihilation rate. The lifetimes of Rydberg Ps states are determined almost entirely by radiative decay and are twice as long as those of the corresponding states in hydrogen atoms~\cite{Bethe1957}. Thus, as long as the scattering cell is sufficiently short that no fluorescence is likely to occur during transit, annihilation events will provide a clear signal of the formation of positron-atom bound states. 
This requires careful control of secondary processes, such as elastic scattering or ionization events that could lead to annihilation following wall collisions.  As we discuss in the Appendix, for the correct choice of experimental parameters the cross sections for these processes compared to those of the interactions of interest can be sufficiently low that the latter will dominate.

\begin{figure}
\includegraphics[width=0.48\textwidth]{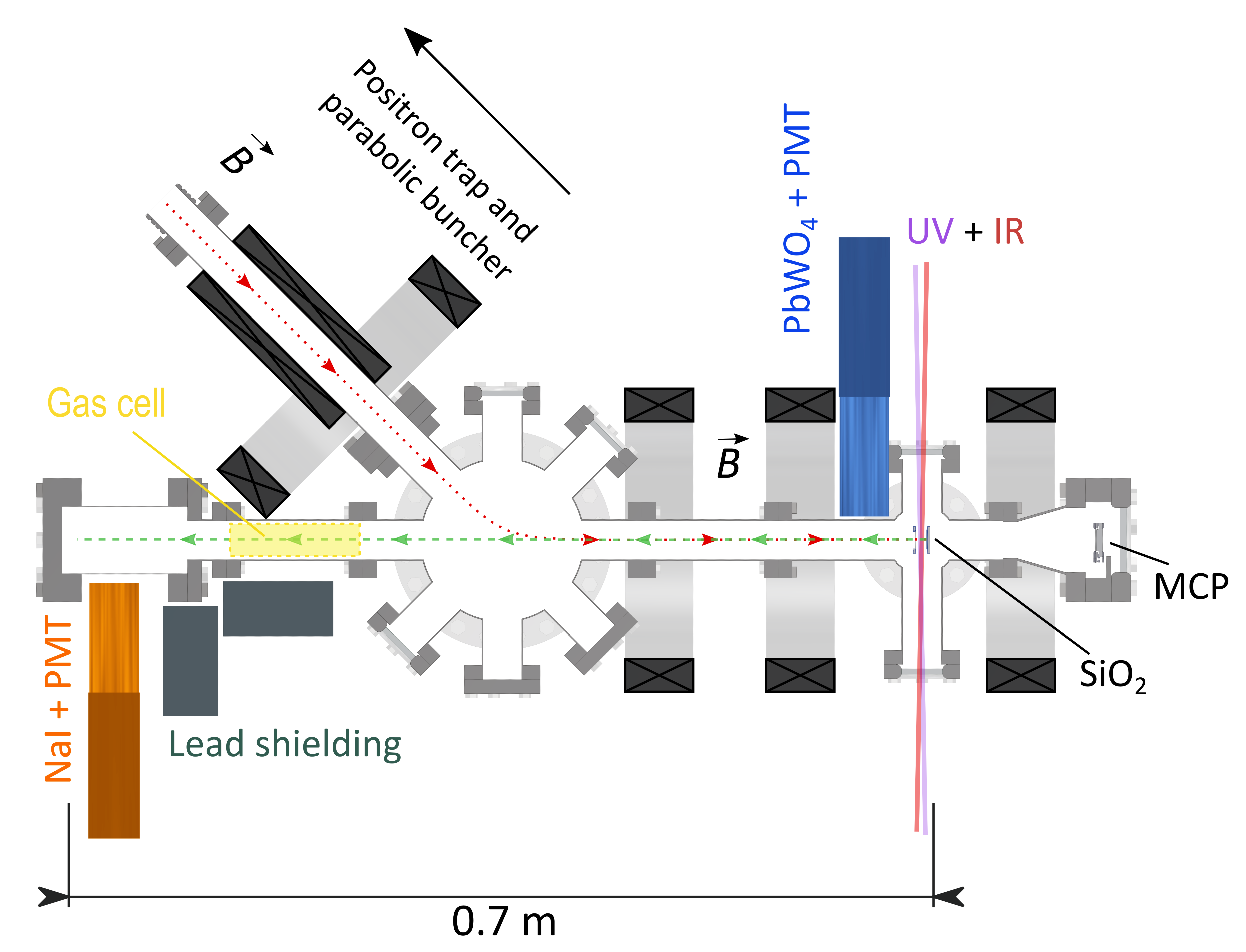}
\caption{\label{fig:TOF} Apparatus for Rydberg Ps production and TOF spectroscopy.  The incident positron beam is guided by the magnetic field of a solenoid and series of four coils (black) through an angle of $45 ^\circ$ to the Ps converter/laser-interaction region.  The dotted red (dashed green) line represent the path of the positrons (Rydberg Ps). The MCP/phosphor screen assembly is used to align the positron beam with the target.}
\end{figure}

Rydberg Ps beams have recently been utilized in Doppler-correction~\cite{Jones14} and time-of-flight (TOF) experiments~\cite{Jones15}. At UCL we have developed a Rydberg Ps beam for fluorescence lifetime measurements and also for the implementation of Ps-atom optics, designed to manipulate the translational motion of Rydberg states using inhomogeneous electric fields~\cite{Cassidy14}. Owing to the manner in which the atoms are created, Ps beams are highly divergent and have correspondingly low transport efficiencies. Focusing such beams with electrostatic lenses is therefore expected to offer significant improvements. Nevertheless, we have been able to produce long-lived Rydberg Ps atoms that traverse a 0.7~m flight path with flight times up to ${\sim}12~\mu$s. A schematic of this arrangement is shown in Fig.~\ref{fig:TOF}.     

The apparatus is an extension of a system designed for laser spectroscopy,
with the same Ps production and excitation methods as described in Ref.~\cite{Cooper2015}. Positrons from a solid-neon moderated~\cite{Mills1986} $^{22}$Na source are captured in a two-stage Surko trap~\cite{Surko1989, Greaves2003} operating at 1~Hz.  The trap output (${\sim} 10^5$~$e^+$ per cycle) is bunched~\cite{Mills1980} and magnetically guided through a 45$^\circ$ turn into the Ps production region (see Fig.~\ref{fig:TOF}). 

The positron beam is implanted into a mesoporous SiO$_2$ film with an energy $E \approx 2$~keV and a time width of $\Delta t \approx 4$~ns. This results in the creation of Ps atoms with kinetic energies of approximately 1~eV, which subsequently cool via collisions with the internal surfaces of the pores before being emitted into vacuum. As a result, the average Ps energy is determined by the incident positron impact energy, until the confinement energy limit is reached, whereupon the Ps energy becomes constant~\cite{Cassidy2010}. Typically Ps is produced in vacuum with an overall efficiency of $\epsilon \sim 0.3 / e^+$~\cite{Liszkay2008a} and longitudinal kinetic energies in the range 50--500~meV~\cite{Deller2015}. The bias applied to an electrode offset 7~mm from the target and orientated parallel to its surface determines the electric field strength in the intervening Ps-laser-interaction region, $\lvert\vec{F}\rvert \sim 0$~V\,cm$^{-1}$~\cite{Alonso16}. For the production of Rydberg Ps it is important to control the electric field in the interaction region as it strongly affects transitions to the Rydberg states.  

The excitation process follows the same two-step scheme used previously~\cite{Ziock90,Cassidy12}, namely, Ps atoms in the $1^3S$ state are driven by UV photons ($\lambda = 243.0$~nm) to the $2^3P_{J}$ level ($J=0$, 1, 2), and a photon in the range of $\lambda = 760$--729 nm (IR) then drives transition to $n \geq 10$. For this system, states up to $n=27$ have been resolved~\cite{Cassidy15}. The production of Ps atoms is monitored via annihilation gamma radiation using a fast PbWO$_4$ scintillator optically coupled to a photomultiplier tube (PMT). This $\gamma$-ray detector and the technique of single-shot positron annihilation lifetime spectroscopy~\cite{Cassidy2006a} can detect changes in the average Ps decay rate in different time windows, and since the decay rates for the Rydberg levels are comparatively small, the excitation of these states can therefore be inferred in this way~\cite{Cooper2015, Cassidy15}. 

Ground-state Ps atoms emitted from a mesoporous silica film will typically travel around 1~cm before annihilating. In order to study Ps interactions with atoms and molecules in a scattering cell it is therefore advantageous to use relatively energetic Ps atoms (e.g., Ref.~\cite{Brawley15}), or to use long-lived Rydberg Ps which can travel much further before radiative decay and subsequent annihilation can occur. In our experiments Ps is detected 0.7~m away from the production region. The probability of any ground-state atoms traveling this far is entirely negligible, and indeed we do not detect any events if the IR laser is off resonance. 

Rydberg Ps atoms arriving downstream are detected using a NaI scintillator, optically coupled to a PMT, located as shown in Fig.~\ref{fig:TOF}. This detector is sensitive to annihilation $\gamma$ rays produced from Ps atoms that (1) were emitted from the film within $1.5^\circ$ of normal to its surface, and (2) have been excited to Rydberg levels able to survive the 2--15~$\mu$s flight time. Considering the solid angle of acceptance (${\sim} 2.3 \times 10^{-3}$~sr) and coverage of the NaI detector (${\sim} 1.7$~sr), the background-subtracted detection rate of 0.02--0.1~Hz equates to  production of roughly (0.2--1.0)${}\times 10^3$ Rydberg Ps per trap cycle, assuming a cosine angular distribution for emission from the film, and neglecting the possibility of in-flight fluorescence or direct annihilation.
\begin{figure}
\includegraphics[width=0.45\textwidth]{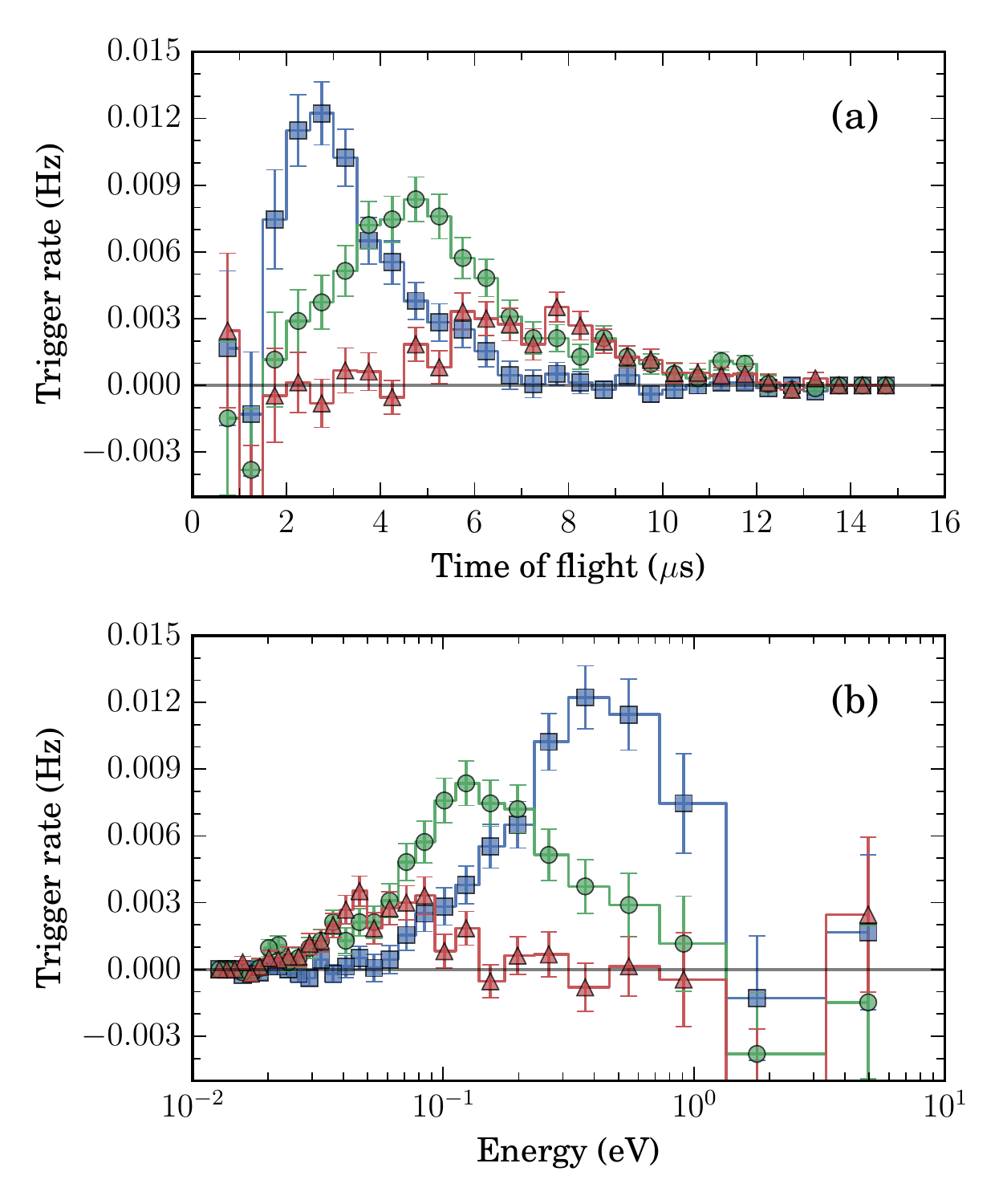}
\caption{\label{fig:NaI} (a)  $n=18$ Ps time-of-flight ($t_z$) as measured along the $z=0.7$~m flight path from the mesoporous SiO$_2$ film to a NaI detector, recorded using 500~ns time bins. (b) the same data as (a) given in terms of the corresponding energy distributions $E_z \approx m_{e}  (z/t_z )^2$. Each series represents a different laser trigger time of 5~ns (\textcolor{bblue}{$\blacksquare$}),  15~ns (\textcolor{ggreen}{$\bullet$}), or 25~ns (\textcolor{rred}{$\blacktriangle$}). }
\end{figure}

Figure~\ref{fig:NaI} shows TOF data recorded using the NaI detector for $n=18$ Ps states.  The lasers were triggered to intersect the excitation region at three different times  relative to the positrons being implanted into the film.  The distribution corresponding to the earliest laser time is the hottest of the three because the Ps atoms that were excited had spent the least amount of time inside of the film before being emitted~\cite{Deller2015}.  The data obtained when the laser was delayed by 10 or 20~ns, however, show colder Ps distributions. This is because the irradiated atoms are those that have had time to cool in the target via inelastic collisions within the pore structure, and also because the fastest atoms have had time to leave the excitation region. 

The data in Fig.~\ref{fig:NaI} illustrate how adjusting the laser delay provides a degree of control over the energy distribution of the Rydberg Ps. It is also possible to control the the Ps energy distribution by changing the positron implantation energy~\cite{Deller2015}. This provides access to a wider energy range but has the disadvantage that this method requires tuning the system in other ways. In general, the positron implantation energy can be used as a gross selector, while the laser delay can provide fine tuning of the Ps energy distribution. The former alters the initial Ps energy distribution according to the cooling in the mesoporous silica, whereas the latter selects different parts of whatever Ps energy distribution is present.   

The TOF spectra shown in Fig.~\ref{fig:NaI} indicate that it is already possible to perform an experiment designed to study the formation of positron-atom bound states. The most direct approach would be would be to insert a gas cell in the path of the Rydberg Ps atoms, as indicated in Fig.~\ref{fig:TOF}. This arrangement could be used to measure the energy thresholds for positron-molecule--bound-state formation due to charge-exchange collisions with Rydberg Ps. Suppression of the above-threshold portions of the TOF spectra would be a clear indicator for such formation and would be highly sensitive to the $n$ state of the incident Ps. The Rydberg states can be chosen from a wide range of possible $n$ values simply by varying the IR laser wavelength~\cite{Cassidy15}, and because the mean energy of the Ps beam can be controlled from a few tens of meV to ${\sim} 1$~eV, a diverse range of molecular species are amenable to study in this way.  

To investigate positron binding to the atoms discussed in Sec.~\ref{sec:numres} (Cu, Mg, and Zn), lower $n$ states would be preferable, in order to observe the energy threshold onset.  Experiments conducted so far have focused on producing Rydberg Ps atoms with $n \geq 10$, but the wavelengths required to populate $n=4$--9 ($\lambda = 972$--767~nm) could be easily achieved using alternative laser dyes. Nonetheless, as the cross section typically varies by over two orders of magnitude for $n=10$--20 (Sec.~\ref{subsec:pred}), predictable attenuation of the Rydberg Ps beam as a function of $n$ would be a strong indication of positron-atom--bound-state formation and could be achieved using our current laser systems. 

The experimental arrangement described here was not originally designed to study positron-atom/molecule bound states, and there are several significant improvements that could be made to optimize the system for these measurements. If the transmitted Rydberg Ps beam is monitored with a microchannel plate (MCP) detector, and the gas cell is observed using a $\gamma$-ray detector, the signal-to-noise ratio would be substantially improved. Furthermore, there are numerous ways in which the gas cell could be located much closer to the target than the arrangement indicated in Fig.~\ref{fig:TOF}, which would allow for significantly higher Rydberg Ps beam intensities. Examples include using a smaller chamber for the positron beam deflection, using a transmission Ps converter (e.g., Ref.~\cite{Andersen2015}), or allowing the incident positron beam to pass through the gas cell in an inline reflection geometry. Such modifications could introduce complications: it might be necessary to collimate the Ps beam, more shielding would be required for the detectors, and the gas cell could potentially cause contamination of the Ps converter, but these would have to be weighed against the corresponding increase in count rates. 

The cross sections of interest span a wide range, and for experimentally accessible parameters are generally quite high, in the range of $10^3$--$10^4 \pi a_{0}^2$ (see Figs. \ref{fig:CS_MgCuZn} and \ref{fig:CS_molecules}). The target gas pressure required to ensure an interaction through a single pass in the scattering cell of length $\ell$ is approximately $1/(\sigma \ell)$. A gas cell 5~cm long would allow for efficient detection, with almost $2\pi$ solid-angle coverage, and the required pressure would then be around $5\times 10^{-5}$ to $5\times 10^{-6}$~Torr. For the molecular target gases (Fig.~\ref{fig:CS_molecules}) this is relatively easy to achieve, but for the metals (Fig.~\ref{fig:CS_MgCuZn}) it is more complicated and requires the use of a heated scattering cell. To obtain a vapor pressure of ${\sim} 10^{-5}$~Torr of Zn or Mg requires heating to 485~K and 555~K, respectively~\cite{Greenbank}, whereas Cu must be heated to around 1200~K~\cite{Geiger1}. These are all experimentally achievable, although since Zn and Mg are considerably easier to implement, these would be the focus of initial studies.  

The basic measurement process relies on the formation of bound states to initiate annihilation events that would not otherwise have occurred. However, competing processes must also be considered, in particular ionization, elastic scattering, and ground-state Ps formation (see Appendix for some ionization cross section estimates). Any of these could provide a signal that would be difficult to distinguish from the events we wish to study. Ionization could be monitored by controlling the electric field in the gas cell. If free positrons are present they can be extracted from the cell, and hence not counted as a spurious signal. If Rydberg Ps atoms undergo elastic scattering or Ps formation they may nevertheless be detected following collisions with the cell or direct annihilation.

Both experiments~\cite{Brawley2010} and calculations~\cite{Fabrikant14} show that ground state Ps atoms scatter from atoms and molecules with total cross sections similar to those of equivelocity electrons. The upper limit for the Rydberg Ps ionization cross section is set by the sum of the equivelocity electron and positron total scattering cross sections, and drops off rapidly close to threshold, as shown in Fig.~\ref{fig:Cu_ioniz}. Thus, we would expect that in the appropriate low-energy range, the bound-state formation cross sections relevant to the proposed experiment may be considerably larger than those of any other process that could give rise to an increased annihilation signal (see Appendix). This will have to be verified by experiment, however, since the calculated cross sections are estimates and, as far the authors are aware, no total cross section data for the particular target atoms to be studied are currently available.

\section{\label{sec:conc}Conclusions}

A new experiment has been proposed to detect the existence of positron-atom bound states. This would be achieved by observing the charge-transfer reaction~(\ref{eq:reac}), with the incident Ps in a Rydberg state. We have provided theoretical estimates for the cross section of this reaction. By comparing these with experimental data, it may be possible to infer a positron-atom binding energy, and compare it with existing theoretical predictions~\cite{Mitroy02,Dzuba12,Harabati14}.

Our calculations were performed in the first Born approximation. The problem was reduced analytically to a one-dimensional integral involving the internal Ps($nl$) wave function in momentum space, and this integral was evaluated numerically. Using the semiclassical approximation, we also obtained a simple analytical expression for the cross section for $l\ll n$. As a check, the method was applied to Ps-H collisions leading to H$^-$, and results were found to be broadly in accord with existing calculations, including the DWBA calculation for $n=1$, performed in this work. We note that the agreement is better for $n=3$, as could be expected for a method that should be valid for higher Rydberg states. Estimates of cross sections were then given for positron binding to Mg, Cu, Zn, $\text{C}_2\text{H}_6$, $\text{CH}_3\text{F}$, and $\text{CH}_3\text{Br}$. In general, the largest cross section in the incident Ps energy range considered (0.001--10~eV) is obtained for $n\sim 1/\sqrt{4\eb}$, i.e., the value of $n$ for which the positron transfer is resonant. At large $n$ and sufficiently large incident Ps energies, the cross section $\sigma$ depends on the Ps principal quantum number $n$ as $\sigma\propto 1/n^3$.

There are some important points to note about our theoretical calculations.
\begin{enumerate}
\item The outgoing electron is treated as a plane wave. To account for the attractive Coulomb field of the positron-atom complex, one can describe the outgoing electron by using a Coulomb wave. This would lead to finite cross sections at threshold energy, but would make the calculation more cumbersome. We expect that the Coulomb interaction would be important only for low outgoing electron energies, $k^2/2 \lesssim 1/2n^2$.
\item The perturbation $V$ in Eq.~(\ref{eq:A_m}) only accounts for the interaction of the positron with the atom; in principle one should also include the interaction of the electron with the atom. However, its effect on the formation of the positron-atom bound state is expected to be comparatively small since Rydberg Ps is a diffuse object. This is also why the role of exchange between the diffuse electron within Ps and strongly bound atomic electrons should be small.
\item The form of the positron wave function used in the derivation is correct for binding by atoms or molecules with ionization potentials $I>6.8$~eV. For atoms with $I<6.8$~eV, the asymptotic wave function corresponds to Ps($1s$) bound to the positive ion, and the dominant form of the bound-state wavefunction is the `Ps-ion cluster'~\cite{Mitroy02}. However, it still contains a `positron-atom component,' and the present cross sections could be used with caution as order-of-magnitude estimates.
\item The presence of electric fields in the experimental setup will lead to Stark mixing of the Ps states. Here we have considered briefly the dependence of the cross sections on the value of $l$ and found that at low Ps energies the cross sections for different $l$ agree to within an order of magnitude. Theoretically, it is feasible to account for the Stark effect rigorously by using the internal Ps wave functions in parabolic coordinates.
\end{enumerate}

We expect that the computed cross sections for the Rydberg-state-Ps--atom collisions are valid to within an order of magnitude or better. Measurements of absolute cross sections would be possible with the molecular targets, all of which are gaseous at room temperature. However, owing to large uncertainties in the target number density in hot cells (e.g., Ref.~\cite{Anderson14}) it is likely that only relative cross sections could be measured for the metal targets. 

Relative cross section measurements could in principle be normalized using accurate calculations, although the applicability of such calculations might be compromised by incomplete knowledge of the Rydberg Stark states produced in the experiment, since they are highly sensitive to stray fields~\cite{Gallagher1994}. Moreover, it is possible that the presence of a background ionized gas of metal atoms in the hot cell will generate large variations in the potential that cannot be controlled or accurately measured, making it impossible to produce well-defined Rydberg Stark states. The extent to which this occurs could be monitored using high-$n$ Ps states, or possibly a secondary beam of Rydberg He atoms, to probe the electric field in the cell~\cite{Osterwalder99}.      

The count rates for our experiments, neglecting improvements obtained by reconfiguring the apparatus (which would likely be substantial), would be the same as those obtained when measuring the TOF distributions shown in Fig.~\ref{fig:NaI}, each of which can be recorded in around 8--10~hours. The measurements would consist of obtaining such spectra for various different conditions (i.e., varying $n$ and the initial velocity distributions) with and without the target gas present. Thus, we would expect to obtain a complete data set sufficient to determine if bound states have been produced (including null tests and verifications) in approximately one week for each target gas. 


\begin{acknowledgments}
The work of A.R.S. has been supported by the Department for Employment and Learning, Northern Ireland. Work at UCL was funded in part by the Leverhulme trust (Grant No. RPG-2013-055), the ERC (Grant No. CIG 630119), and the EPSRC (Grant No. EP/K028774/1). We gratefully acknowledge assistance from A. M. Alonso and B. S. Cooper in operating and maintaining the experimental apparatus. 
\end{acknowledgments}

\appendix*

\section{Ps ionization in collisions with atoms}

Consider the process of ionization of Rydberg Ps in a collision with a target atom/molecule $A$,
\begin{equation}\label{eq:breakup}
\text{Ps}(nl) + A \longrightarrow A+e^-+e^+.
\end{equation}
Because of the small binding energy of $\text{Ps}(nl)$, its constituent electron and positron can be considered as quasifree during their interaction with the target. This allows one to use the impulse approximation (IA) (see, e.g., Refs.~\cite{Fabrikant14,Chew52,Matsuzawa84,Fabrikant92,Starrett05})
and write the corresponding amplitude in the form
\begin{align*}
B_m=-2\pi\Bigg[&\int \widetilde{\psi}_f^*\left( \vec{q} + \frac{\Delta\vec{K}}{2} \right) f_e(\vec{k}_e',\vec{k}_e) \widetilde{\psi}_{nlm}(\vec{q}) \frac{d^3\vec{q}}{(2\pi)^3} \nonumber\\
{}+{}&\int \widetilde{\psi}_f^*\left( \vec{q} - \frac{\Delta\vec{K}}{2} \right) f_p(\vec{k}_p',\vec{k}_p) \widetilde{\psi}_{nlm}(\vec{q}) \frac{d^3\vec{q}}{(2\pi)^3}\Bigg],
\end{align*}
where $\widetilde{\psi}_{nlm}$ and $\widetilde{\psi}_f$ are the internal wave functions of the incident and final-state Ps in momentum space, $f_e$ ($f_p$) is the electron (positron) scattering amplitude from the target with initial and final momenta $\vec{k}_{e,p}=\vec{K}/2\pm \vec{q}$ and
$\vec{k}_{e,p}'=\vec{K}/2\pm \vec{q}+\Delta\vec{K}$, respectively, and 
$\Delta\vec{K}=\vec{K}'-\vec{K}$ is the difference between the final
and initial Ps center-of-mass momenta $\vec{K}'$ and $\vec{K}$.

The typical electron and positron momenta within $\text{Ps}(nl)$ are small,
${\sim} 1/n$. This means that for small incident Ps momenta $K\ll 1$~a.u.,
the initial and final electron and positron momenta in the amplitudes
$f_e$ and $f_p$ are also small. Hence, we can approximate these amplitudes by their $s$-wave contributions and take them in the limit $k_{e,p}\rightarrow 0$ for the simplest estimate:
\begin{subequations}
\begin{align}
f_e(\vec{k}_e',\vec{k}_e) \simeq -a_-,\\
f_p(\vec{k}_p',\vec{k}_p) \simeq -a_+,
\end{align}
\end{subequations}
where $a_-$  and $a_+$ are the $e^-$-$A$ and $e^+$-$A$ scattering lengths~\cite{LandauQM}, respectively. Then
\begin{align}
B_m=2\pi\Bigg[&a_- \int \widetilde{\psi}_f^*\left( \vec{q} + \frac{\Delta\vec{K}}{2} \right)  \widetilde{\psi}_{nlm}(\vec{q}) \frac{d^3\vec{q}}{(2\pi)^3} \nonumber\\
{}+{}&a_+ \int \widetilde{\psi}_f^*\left( \vec{q} - \frac{\Delta\vec{K}}{2} \right)  \widetilde{\psi}_{nlm}(\vec{q}) \frac{d^3\vec{q}}{(2\pi)^3}\Bigg].
\end{align}

Let $\vec{k}$ be the internal momentum of Ps after the ionization. Neglecting the Coulomb interaction between the electron and positron in the final state, we write the internal wave function as a plane wave: $\psi_f(\vec{r}_1-\vec{r}_2)=\exp[i\vec{k}\cdot(\vec{r}_1-\vec{r}_2)]$. Then $\widetilde{\psi}_f(\vec{s})=(2\pi)^3\delta(\vec{s}-\vec{k})$, and we obtain
\begin{align*}
B_m=2\pi \left[ a_- \widetilde{\psi}_{nlm}\left( \vec{k}-\frac{\Delta\vec{K}}{2} \right) + a_+ \widetilde{\psi}_{nlm}\left( \vec{k}+\frac{\Delta\vec{K}}{2}\right)\right].
\end{align*}
Instead of using the final-state momenta $\vec{K}'$ and $\vec{k}$, let us use the final electron and positron momenta $\vec{k}_1$ and $\vec{k}_2$,
respectively. Then $\displaystyle\vec{k}_1=\vec{K}'/2+\vec{k}$ and $\displaystyle\vec{k}_2=\vec{K}'/2-\vec{k}$, giving a more convenient form for the amplitude:
\begin{align*}
B_m=2\pi \left[ a_- \widetilde{\psi}_{nlm}\left( \frac{\vec{K}}{2}-\vec{k}_2 \right) + a_+ \widetilde{\psi}_{nlm}\left( \vec{k}_1-\frac{\vec{K}}{2}\right)\right].
\end{align*}

\begin{widetext}
The $m$-dependent ionization cross section $\sigma_m$ is found from
\begin{align}
d\sigma_m(\vec{K})=\frac{2\pi}{K/2} \left\lvert B_m \right\rvert^2 \delta\left( \frac{k_1^2}{2} + \frac{k_2^2}{2}-\frac{K^2}{4} - \frac{1}{4n^2} \right) \frac{d^3\vec{k}_1}{(2\pi)^3} \frac{d^3\vec{k}_2}{(2\pi)^3}
\end{align}
Writing $k_2^2\, dk_2=k_2\, d(k_2^2/2)$ and integrating over $d(k_2^2/2)$ yields the triple differential cross section:
\begin{align}\label{eqn:tdcs1}
\frac{d^3\sigma_m}{dk_1\, d\Omega_{\vec{k}_1}\, d\Omega_{\vec{k}_2}}=\frac{2 k_1^2 k_2}{(2\pi)^3 K}
\left\lvert a_- \widetilde{\psi}_{nlm}\left( \frac{\vec{K}}{2}-\vec{k}_2 \right) + a_+ \widetilde{\psi}_{nlm}\left( \vec{k}_1-\frac{\vec{K}}{2}\right) \right\rvert^2,
\end{align}
with the energy conservation law
\begin{equation}
k_2 = \sqrt{\frac{K^2}{2} - \frac{1}{2n^2} - k_1^2} .   
\end{equation}
Separating the momentum-space wave functions into radial and angular parts [see Eq.~(\ref{eqn:FYprod})] and averaging the cross section over the magnetic quantum number $m$ of the incoming Ps, we find
\begin{align}\label{eqn:TDCS}
\frac{d^3\sigma}{dk_1\,d\Omega_{\vec{k}_1}\,d\Omega_{\vec{k}_2}}
=
\frac{k_1^2 k_2}{2\pi K}
\Bigg\{
&a_-^2 \big\lvert F_{nl}(\lvert\vec{K}/2-\vec{k}_2\rvert)\big\rvert^2
+
a_+^2 \big\lvert F_{nl}(\lvert\vec{k}_1-\vec{K}/2\rvert) \big\rvert^2 \nonumber\\
{}+{}&
2 a_- a_+ F_{nl} (\lvert\vec{K}/2-\vec{k}_2\rvert) F_{nl}(\lvert\vec{k}_1-\vec{K}/2\rvert)
P_l\left[ \frac{(\vec{K}/2-\vec{k}_2)\cdot(\vec{k}_1-\vec{K}/2)}{\lvert \vec{K}/2-\vec{k}_2\rvert \lvert \vec{k}_1-\vec{K}/2 \rvert} \right]
\Bigg\},
\end{align}
where $P_l$ is the Legendre polynomial.
\end{widetext}

As the Rydberg Ps in the experiment described in Sec.~\ref{sec:exp} is produced mainly in $s$ and $d$ states, we consider the cases $l=0$ and $l=2$. Integrating the differential cross section~(\ref{eqn:TDCS}) and
using the same variable substitution that led to Eq.~(\ref{eq:cs3}),
we find the total ionization cross section for $l=0$ as
\begin{align}
\sigma&=\frac{2\pi k_{\max}^4}{K}\Bigg[ \left(a_-^2+a_+^2\right) \int_0^{\pi /2}I_2(k_\text{max}\sin \alpha )\sin ^22\alpha \,d\alpha \nonumber \\
&{}+ 2a_-a_+  \int_0^{\pi /2}I_1(k_\text{max}\sin \alpha ) I_1(k_\text{max}\cos\alpha ) \sin ^22\alpha \,d\alpha \Bigg] ,\label{eqn:ion_cs_l0}
\end{align}
where $k_\text{max}=\sqrt{K^2/2-1/2n^2}$, the variable $\alpha $ determines the partition of the total kinetic energy between the electron and the positron ($k_1=k_{\rm max}\sin \alpha $, $k_2=k_{\rm max}\cos \alpha $), and 
\begin{subequations}
\begin{align}
I_1(k) &= \frac{1}{kK}\int_{\lvert k-K/2\rvert}^{k+K/2} F_{n0}(p) p\, dp, \label{eq:I1}\\
I_2(k) &= \frac{1}{kK}\int_{\lvert k-K/2\rvert}^{k+K/2} \left\lvert F_{n0}(p)\right\rvert^2 p\, dp. \label{eq:I2}
\end{align}
\end{subequations}
As the impulse approximation is valid in the limit of large $n$ and low $K$, the cross section was computed numerically for $n=5$--20 with incident Ps energies ${\leq}1$~eV. It was found that the contribution of the interference term [the second term in square brackets in Eq.~(\ref{eqn:ion_cs_l0})] is negligible. This is caused by the oscillatory behaviour of $F_{n0}(p)$, which suppresses the integral $I_1(k)$, Eq.~(\ref{eq:I1}). Hence, we calculated the total cross section for $l=2$
from Eq.~(\ref{eqn:ion_cs_l0}) without the interference term, using
$F_{n2}$ instead of $F_{n0}$ in Eq.~(\ref{eq:I2}). Neglecting the interference term also allows one to derive the ionization cross section in the semiclassical approximation, by using the classical momentum distribution~(\ref{eq:w_n}) instead of $\lvert F_{nl}(p)\rvert^2p^2$ for $l\ll n$
(cf.~Sec.~\ref{subsec:sc}). 

The positron-atom scattering length $a_+$ can be estimated from the known binding energy through $a_+\approx -1/\sqrt{2\eb}$~\cite{LandauQM}. The $e^-$-$A$ scattering length can similarly be estimated from the target's electron affinity (EA)~\footnote{For an atom with a single valence electron, such as Cu, the electron affinity determines the scattering length for the Ps-atom collision in which the total electron spin $S=0$. The magnitude of the scattering length for $S=1$ is expected to be smaller.}. Since the positron binding energy is usually small, the positron contribution to the ionization cross section dominates.

\begin{figure}
\includegraphics{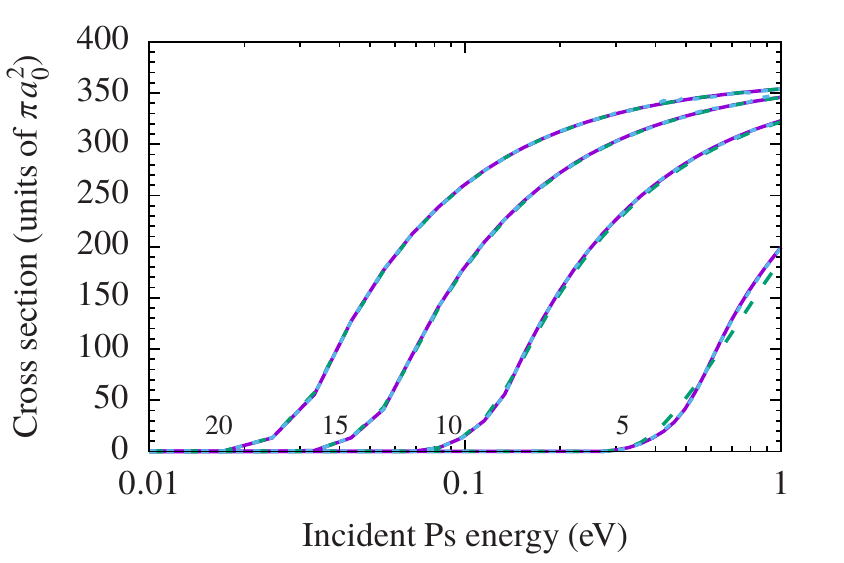}
\caption{Ionization cross sections for $\text{Ps}(nl)$ collisions with Cu: solid curves, $l=0$ [Eq.~(\ref{eqn:ion_cs_l0})]; dashed curves, $l=2$ [Eq.~(\ref{eqn:ion_cs_l0}) without the interference term and with $F_{n2}$ used in (\ref{eq:I2}) instead of $F_{n0}$]; dotted curves, results obtained from Eq.~(\ref{eqn:ion_cs_l0}) neglecting the interference term and using the classical momentum distribution (\ref{eq:w_n}) instead of $\lvert F_{n0}(p)\rvert^2 p^2$.  The values of $n$ are shown next to the curves.}
\label{fig:Cu_ioniz}
\end{figure}

As an example, Fig.~\ref{fig:Cu_ioniz} shows ionization cross sections for Ps collisions with the Cu atom for $l=0$ (with interference) and $l=2$ (without interference), as well as the semiclassical result. We take the electron affinity to be 1.235~eV~\cite{CRC86}. The cross sections for $l=0$ and $l=2$ are almost indistinguishable, except for the lowest principle quantum number $n=5$. In the scale of Fig.~\ref{fig:Cu_ioniz}, the semiclassical cross section for $l=0$ is identical to the quantum calculation, which confirms that the interference term is negligible.

The cross sections grow monotonically from zero at threshold, and in the limit of large Ps energy they become constant. The value of the cross section at large $K$ (though still ${\ll} 1$~a.u.) may be determined as follows. The typical electron and positron momenta in the Ps Rydberg state $nlm$ are small (${\sim}1/n$), so for $K\gg 1/n$ one can replace the corresponding momentum-space densities by the delta functions. Neglecting interference, Eq.~(\ref{eqn:tdcs1}) gives
\begin{equation*}
\frac{d^3\sigma}{dk_1\,d\Omega_{\vec{k}_1}\,d\Omega_{\vec{k}_2}}=\frac{2k_1^2k_2}{K} \left[ a_-^2 \delta\left(  \frac{\vec{K}}{2}-\vec{k}_2\right) + a_+^2 \delta\left(  \vec{k}_1-\frac{\vec{K}}{2}\right) \right],   
\end{equation*}
which yields
\begin{equation}\label{eqn:cs_highK}
\sigma=4\pi\left(a_-^2 + a_+^2\right).    
\end{equation}
This result arises because the electron and positron in the incident weakly bound Ps are quasifree, each with momentum $K/2$. The total ionization cross section is then simply the sum of the electron-atom and positron-atom (elastic) scattering cross sections, $\sigma =\sigma _-+\sigma _+$. Unlike Eq.~(\ref{eqn:cs_highK}), this latter result is valid for any Ps momentum $K$. Instead of having a plateau, the ionization cross section will then decrease with the Ps energy, following the decrease of the positron and electron cross sections $\sigma _{\pm}$.

For Cu, supposing the incident Ps has an energy of 0.05~eV, for $n=15$ the ionization cross section is estimated to be about $25\pi a_0^2$, while the charge-transfer cross section is approximately $125\pi a_0^2$, i.e., much greater. Depending on the Ps energy and the value of $n$, this may or may not be the case, but we have shown that there should be a `window' of Ps energies and values of $n$ where charge transfer is the dominant process. In particular, it appears that for the Ps principal quantum numbers for which the charge-transfer cross section is largest, it is also much greater than the corresponding ionization cross section, making the proposed detection scheme feasible.


%

\end{document}